\documentclass[fleqn,usenatbib,linenumbers,twocolumn]{mnras}

\usepackage[T1]{fontenc}
\usepackage{ae,aecompl}
\usepackage{graphicx}	
\usepackage{amsmath}	
\usepackage{amssymb}	
\usepackage{subfigure}
\usepackage{threeparttable}
\usepackage{mathrsfs}

\usepackage{CJKutf8}
\usepackage{graphicx}
\usepackage{amsmath}
\usepackage{amssymb}
\usepackage{cases}
\usepackage{mathrsfs}
\usepackage{pifont}
\usepackage{amsmath}
\usepackage{epstopdf}
\usepackage{xcolor}
\usepackage[all]{hypcap}
\usepackage{mwe,tikz}
\usepackage{bm}
\usepackage{tabularx}
\usepackage{subfigure}
\usepackage{xspace}
\usepackage{rotating}
\usepackage{verbatim}
\usepackage{appendix}
\usepackage[utf8]{inputenc}
\usepackage{cleveref}

\title{A comptonized fireball bubble: physical origin of magnetar giant flares}

\author[Zhang et al.]{
		Zhao Joseph Zhang$^{1,2}$\thanks{E-mail: zhangzhao981013@outlook.com}, Bin-Bin Zhang$^{1,3,4}$\thanks{E-mail: bbzhang@nju.edu.cn}, Yan-Zhi Meng$^{1,4}$\thanks{E-mail: yzmeng@nju.edu.cn}
			\\
			$^{1}$School of Astronomy and Space Science, Nanjing University, Nanjing 210023, China\\
    $^{2}$Department of Earth and Space Science, Osaka University, 1-1 Machikaneyama, Toyonaka, Osaka 560-0043, Japan\\ $^{3}$Purple Mountain Observatory, Chinese Academy of Sciences, Nanjing 210023, China\\			
   $^{4}$Key Laboratory of Modern Astronomy and Astrophysics (Nanjing University), Ministry of Education, Nanjing 210023, China\\
		}
\date{Accepted 2023 February 6. Received 2023 February 4; in original form 2022 September 29
}
\pubyear{2023}

\begin{document}
\label{firstpage}
\pagerange{\pageref{firstpage}--\pageref{lastpage}}
\maketitle
		
\begin{abstract}	
Magnetar giant flares (MGFs) have been long proposed to contribute at least a sub-sample of the observed short gamma-ray bursts (GRBs). The recent discovery of the short GRB 200415A in the nearby galaxy NGC 253 established a textbook-version connection between these two phenomena. Unlike previous observations of the Galactic MGFs, the unsaturated instrument spectra of GRB 200415A provide for the first time an opportunity to test the theoretical models with the observed $\gamma$-ray photons. This paper proposed a new readily fit-able model for the MGFs, which invokes an expanding fireball Comptonized by the relativistic magnetar wind at photosphere radius. In this model, a large amount of energy is released from the magnetar crust due to the magnetic reconnection or the starquakes of the star surface and is injected into confined field lines, forming a trapped fireball bubble. After breaking through the shackles and expanding to the photospheric radius, the thermal photons of the fireball are eventually Comptonized by the relativistic $e^{\pm}$ pairs in the magnetar wind region, which produces additional higher-energy gamma-ray emission. The model predicts a modified thermal-like spectrum characterized by a low-energy component in the Rayleigh-Jeans regime, a smooth component affected by coherent Compton scattering (CC) in the intermediate energy range, and a high-energy tail due to the inverse Compton process. By performing a Monte-Carlo fit to the observational spectra of GRB 200415A, we found that the observation of the burst is entirely consistent with our model predictions. 
\end{abstract}
		

\section{introduction}
 
Magnetars \citep{Mereghetti2015,Kaspi2017} belong to a particular type of neutron stars (NS) carrying ultra-strong magnetic field typically in the range of B $\sim10^{14}$-$10^{15}$ G \citep{TD1992}, much higher than the critical magnetic field, $B_Q = 4.4\times10^{13}$ G, above which the nonrelativistic Landau energy becomes comparable to the electron rest energy \citep{TD1995}. Observationally, magnetars are registered as two classes: the Anomalous X-ray Pulsars (AXPs) and the soft gamma-ray repeaters (SGRs). SGRs have recently aroused increasing research interest due to their associations with fast radio bursts \citep[FRBs;][]{Ioka2016,Metzger2019,Katz2020,Linlin2020,2020Natur.587...45Z,Xiao2021} and gamma-ray bursts \citep[GRBs;][]{Hurley1999, YJ2020,RobertNA2021,Svinkin2021Na}. In particular, the magnetar giant flares (hereafter, MGFs) from the SGRs \citep{Mazets1979a, Mazets1979b} have been long considered to be a subclass of short GRBs \citep{Laros1986,Atteia1987}. 

 The first magnetar giant flare (GRB 790305B) was observed from SGR 0526-66 \citep{Golenteskii1984}, which is located in the star-forming Dorado region in the large magellanic cloud (LMC). Since then, two more Galactic MGF events have been confirmed, namely, the MGF of SGR 1900+14 on 1998 August 27 \citep{Feroci2002} and the MGF of SGR 1806-20 on 2004 December 27 \citep{Yamazaki2006}. Located within a few $\sim$ kpc, those events, although having provided enriched data in studying their temporal features (including the short abrupt rise, the quasi-exponential decay, the subsequent pulsating tail; \citealt{Hurley1999}; and the quasi-periodic oscillations; \citealt{Israel2005, Strohmayer2005}) of the MGFs, are all observed saturated by the $\gamma$-ray detectors due to their overwhelmingly large numbers of photons. Thus, studying the accurate spectral data of MGF has been infeasible until the extragalactic MGF-originated event, GRB 200415A, was observed.

 GRB 200415A is an apparent short GRB discovered in the nearby Sculptor galaxy (NGC 253), which is located about 3.5 Mpc away \citep{Bissaldi2020}, much further than those in the previous MGF sample. A comprehensive analysis performed by \cite{YJ2020} suggests the burst is a significant outlier of both the Type I and Type II GRBs, but otherwise entirely consistent with being an MGF event in terms of the temporal and spectral features. Thanks to its significant distance outside the Milky Way, the bright GRB 200415A is not subject to instrumental saturation and provides an ideal case to study the photon behaviors of the MGFs. A spectral fit using some empirical functions suggested that the burst is dominated by thermal-like emissions \citep{YJ2020}, likely originates from an expanding fireball outside the magnetar surface and may be Doppler-boosted by the relativistic wind \citep{RobertNA2021}. 
 
 Nevertheless, as a rare and most powerful type of SGR activity, MGFs' physical origin, trigger mechanism, and radiation process are still not fully unveiled. A trigger of an MGF can be caused either internally \citep[e.g. large-scale star crust fracturing caused by the shear forces against the motion of the magnetic footpoints; neutral point reconnection due to the torsional of the twisted interior magnetic field;][]{Parker1983a,Parker1983b,TD1995} or externally \citep[e.g., interchange instability and/or magnetic reconnection;][]{Moffatt1985,TD2001}. On the other hand, the radiation process models of MGFs are relatively less diverse. \citet{TD1995} proposed a ``trapped fireball" model to explain the March 5th event on SGR 0526-66. The model introduces an initially expanding fireball to explain the first sharp spike of the MGF event. The fireball is eventually trapped by optically thick pair plasma in the stellar magnetosphere, producing repeated pulsations. In addition, \citet{TD1995} suggested that a relativistic wind driven by the pressure of the electron-positron pair ($e^{\pm}$ hereafter) plasma from the NS surface, which can also release a fraction of the star's energy. The prediction of the E$_{\rm p}$-flux correlation of such a model was claimed to be consistent with the observation of GRB 200415A \citep[][]{RobertNA2021}. Recently, \cite{Van2013, Van2016} and \cite{Barchas2021} conducted some detailed studies of polarized radiation transfer of photons in two modes (E-model: perpendicular to the magnetic field and O-mode: parallel to the field) in the highly magnetized surface locales of magnetar and explored the effect on its scattering opacity. Furthermore, \cite{Van2016} also proposed a model for MGFs that a relativistic radiation-driven outflow along a narrow open field-line bundle. These studies may shed light on the origin and radiation mechanisms of the MGFs. Nevertheless, a detailed first-principle calculation of the MGF models and their validation through a direct fit to the observed spectra is still missing to date. 
 
 Motivated by the previous studies, in this paper, we studied in detail how the fireball expands from the NS surface and penetrates the area of the magnetar wind. Our finding suggests that the fireball is eventually Comptonized by the dense $e^{\pm}$ pair plasma at a much larger radius and produces a multicomponent thermal-like radiation spectrum. Moreover, we directly compare this model with the spectral data of GRB 200415A. This paper is organized as follows. We describe the physical picture of our model in Section \ref{sec:physical_picture}. In Section \ref{sec:RF}, we formulized the radiation mechanism for the calculation of the specific flux and fit our model to the spectral data, constraining some physical parameters. A brief summary is presented in Section \ref{sec:summary}.

	\section{The physical picture} \label{sec:physical_picture}

Our model requires a magnetar characterized by a large-scale dipolar magnetic field with some small-scale and non-axisymmetric magnetic topology \citep{Gourgou2016}. The small-scale field can be caused by internal motions such as the hall drift of the crustal magnetic field. As illustrated in Fig. \ref{fig:Magnetic_structure}, the local small-scale magnetic field lines will be strongly wound up so that the toroidal component is greater than the poloidal dipole strength. Therefore, the local magnetic field may be an order of magnitude higher than that of the large-scale dipolar field, which can exceed $10^{16}$ G. Under such an assumption, the physical picture of our model can be outlined in the following steps:

\begin{figure}
 \label{fig:Magnetic_structure}
 \centering
 \includegraphics[width=0.47\textwidth]{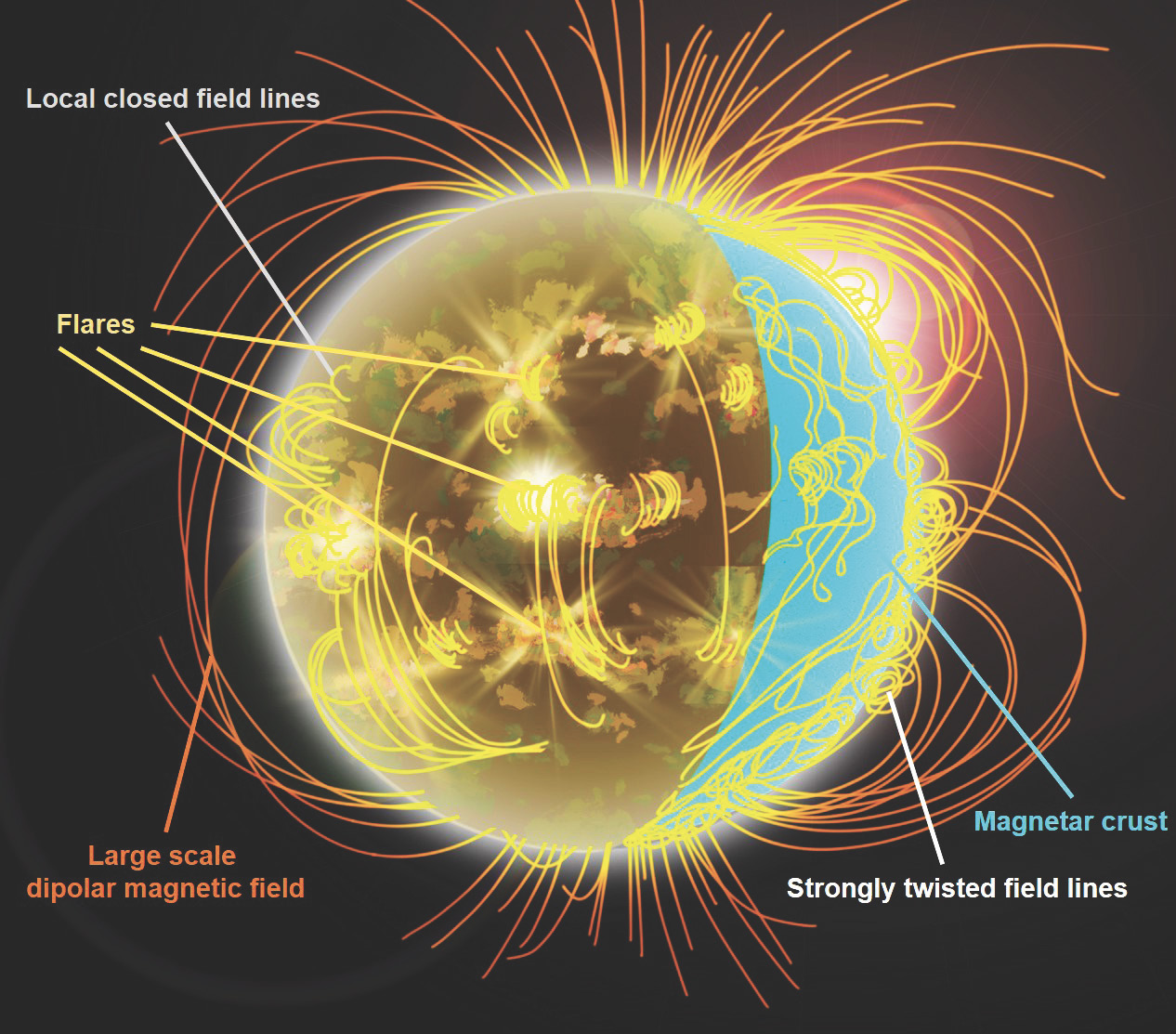}
 \caption{An illustration of the magnetic topology of the magnetar in our model.}
\end{figure}

\begin{enumerate}
\item The occurrence of the small-scale magnetic instability. The Hall drift of the interior field lines and the activity of the NS crust can make the local external field become extremely unstable, causing strong magnetic reconnection.

\item The formation of the trapped fireball. 
Near the NS surface, an enormous amount of energy carried by photons is instantly released via the small-scale magnetic reconnection and interchange instability, and numerous $e^{\pm}$ pairs escaped from the NS due to the fracturing of the crust. These photons and pairs, coupling with each other, are injected into the magnetosphere. A fraction of the energy of the photon-rich pair plasma is confined by the closed magnetic field lines and forms a trapped fireball as shown in Fig. \ref{fig:GFModel}. 

Within the trapped fireball, the energy is gradually dissipated via thermal radiation with a time-scale of up to hundreds of seconds. A pulsating tail is expected within such a long times-cale due to the continuous local quasi-periodic activities of the magnetar. The total energy carried by the pulsating tail is roughly estimated as $E_{\rm tail}\sim 10^{44}$ ergs in the previous three MGFs \citep{Mereghetti2008}. To confine this amount of energy in the closed field lines, the pressure at the outer boundary of the field loop is required to satisfy\citep[][]{Yang2015}
\begin{equation}
\frac{B^2_{R_{s}+l_0}}{8\pi}\gtrsim \frac{E_\mathrm{tail}}{3l_0^3} ,\label{eq1}
\end{equation} 
where $R_s$ is the surface radius of the magnetar, and $l_0$ is the scale of the trapped fireball which in this case is smaller than $R_s$, i .e., $l_0 < R_{s}$ \citep[c.f.,][]{Boggs_2007} . $B_{r}$ is defined by $ B_{r}= B_{*}(r/R_{s})^{-3}$ for a dipolar field at radius of $r$, where $B_{*}$ is the characteristic surface magnetic field of the neutron star.

Observationally, the size of the fireball can be estimated by 
\begin{equation}
\begin{split}
l_0 \sim \Big(\frac{L_\mathrm{\rm tail}}{2\pi caT_\mathrm{tail}^4}\Big)^{1/2} \sim & 8\times10^{4}\ {\rm cm} \bigg(\frac{L_\mathrm{tail}}{ 6 \times 10^{43} \mathrm{\ erg\ s^{-1}}}\bigg)^{1/2} \\ 
&\times \bigg(\frac{kT_\mathrm{tail}}{140 \ \mathrm{keV}} \bigg)^{-2}, \label{eq:l0}
\end{split}
\end{equation}

where $a$ is the radiation constant, and $T_{\rm tail}$ is the thermal equilibrium temperature which can be estimated by the cutoff energy $T_\mathrm{cut}$. The characteristic values of $L_\mathrm{tail} \sim 2 \times 10^{43} {\rm erg}\ {\rm s}^{-1}$ in equation (\ref{eq:l0}) are comparable to those observed in SGR 1806-20, and $kT_{\rm tail} = 140 $ keV is obtained from the spectral fitting of blackbody model for the tail phase \citep{YJ2020}.

\begin{figure*}
 \label{fig:GFModel}
 \begin{center}
 \includegraphics[width=0.9\textwidth]{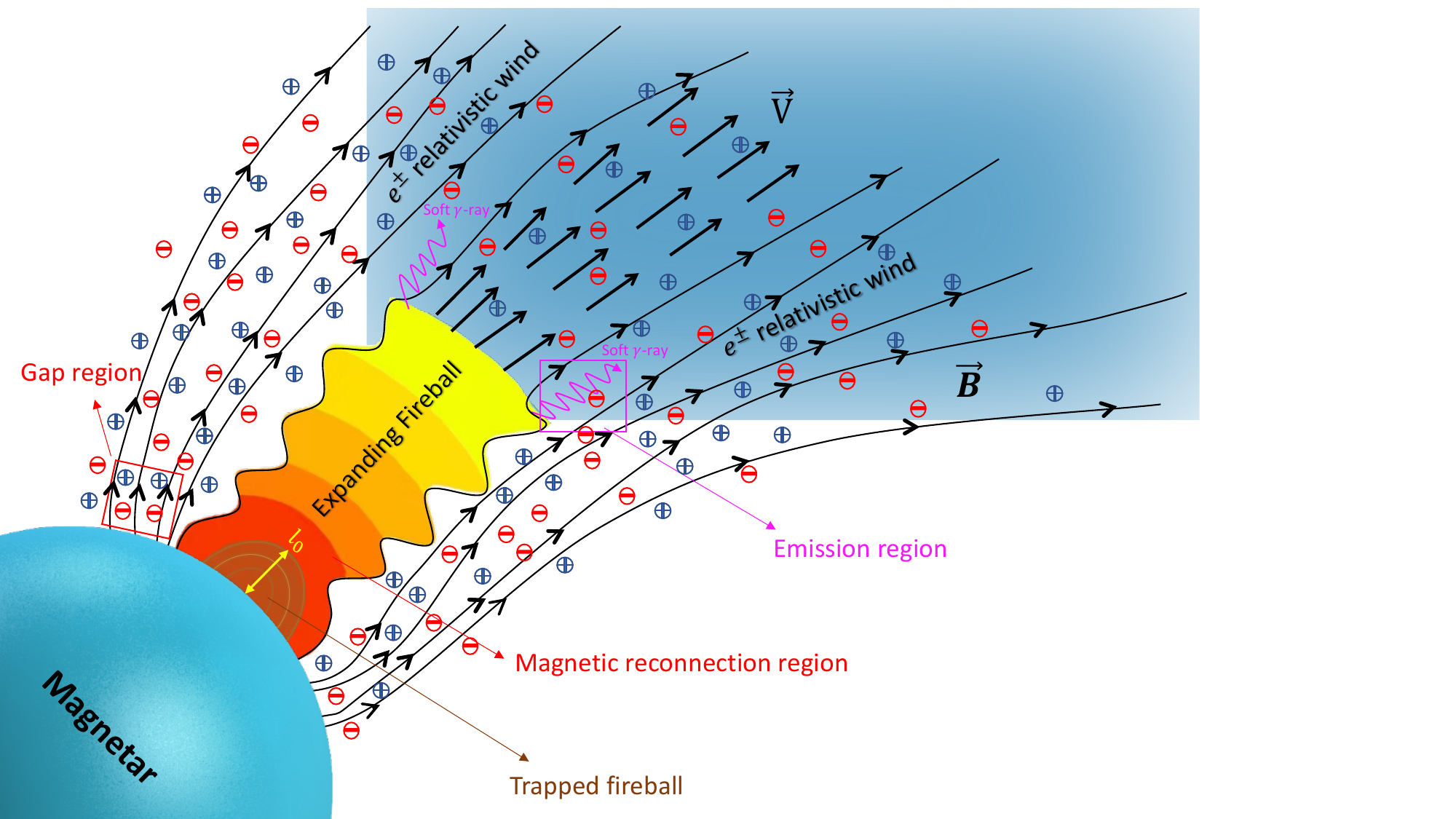}
 \end{center}
 \caption{A schematic diagram of our model. A large number of high-energy photons accompanied by electron-positron pairs are released near the NS surface. A fraction of them is trapped in the confined magnetic field lines (dark red region). The other part breaks through the magnetic field lines due to sufficient radiation pressure and spreads out in the form of a relativistic expanding fireball along the magnetic tube. A large amount of dense $e^{\pm}$ pairs (marked red and blue circle with $\pm$ signs) are emitted from the gap region (red square), and accelerated by the gap potential difference. The pairs further propagate along the open field lines and form a wind zone at a larger radius (open blue region). The expanding fireball eventually penetrates the wind area, and the photons are scattered by the massive pairs in the emission region at photosphere radius (purple square), producing the high-energy $\gamma$-ray photons.}
\end{figure*}

\item The formation of the magnetar wind. 
Due to the significant pressure of the photon-pair plasma, numerous $e^{\pm}$ is driven from the magnetosphere. These pairs, after being accelerated by the gap potential difference, produce the high-energy curvature radiation, which is influenced by the magnetic field, in turn, can be converted into secondary $e^{\pm}$ pairs. As a result, the $e^{\pm}$ pairs increase rapidly in number and move along the field lines, forming a relativistic wind (Fig. \ref{fig:GFModel}). The magnetar wind will supply a high number density of charged particles (i.e., electrons and positrons) which can be calculated \citep{Kumar2020} as a function of the magnetic field, $B_*$, the rotation period, $P$, and the distance to the center of the magnetar, $R$, as

\begin{equation}
\begin{split}
&n_{\pm} = \frac{\mathcal{M} \boldsymbol{B_0} \cdot \Omega_*}{2\pi q_e c}\\ &\approx 4\times10^{23} \ \mathrm{cm}^{-3}\bigg(\frac{\mathcal{M}}{10^8}\bigg)\bigg(\frac{P}{1\ s}\bigg)^{-1}\bigg(\frac{B_{\mathrm{*}}}{10^{16}\ \mathrm{G}}\bigg)\\
&\quad \times \bigg(\frac{R}{R_\mathrm{s}}\bigg)^{-3}, \label{eq:num}
\end{split}
\end{equation}
where $q_e$ is the amount of charge of the electron and $\mathcal{M}$ is the multiplicity parameter defined as the ratio of maximum Lorentz factor of primary electrons and secondary $e^\pm$ pairs. The exact value of $\mathcal{M}$ is highly uncertain, which is related to the magnetic field, the height of the acceleration region, and inversely proportional to the rotation period. For pulsar with magnetic field strength around $10^{12}$ G, $\mathcal{M}$ can be up to $10^6$, and for a magnetar with field strength close to $10^{16}$ G in this study, we estimate $\mathcal{M}$ to be $10^8$ \citep{Ruderman1975}.

\item The expanding of the fireball. 
The fireball will break through the shackles of the field lines and expand along the magnetic tube when the radiation pressure of the fireball is stronger than the magnetic pressure. The local magnetic field lines become open during the expansion (Fig. \ref{fig:GFModel}). For a dipolar field, the bulk Lorentz factor and comoving temperature are determined by the local acceleration as well as the size of expanding fireball, $l$:
\begin{equation}
 \Gamma = (l/l_0)^{3/2},\quad T = T_{\rm ini}(l/l_0)^{-3/2}, \label{eq:Gamma_T}\\
\end{equation} 
where $T_{\rm ini}$ and $l_0$ are the initial temperature and radius of the trapped fireball, respectively. We set $kT_{\rm ini} = 598.6$ keV, which is obtained from observations of the GRB 200415A \citep{YJ2020}.

\item The interaction between the expanding fireball and the wind. The fireball expands relativistically and penetrates to the wind zone within the magnetosphere. 
The photosphere radiation is generally considered to be thermal; therefore, blackbody radiation is often used to describe the spectrum. However, since an extraordinarily high density of $e^{\pm}$ pairs with a thermal distribution is supplied by the magnetar wind, an extra finite thermal medium (Compton cloud) emerges in the front of the expanding path of the fireball, which is highly opaque and leads to a series of radiation transfer effects including scatterings and absorption \citep{Rybicki1986, Beloboro2010, RobertNA2021}. The occurrence of these energy transfer effects will change the spectral shape significantly.
The photons, from either the thermal radiation of the fireball or the annihilation of $e^\pm$ in the wind, are continuously scattered by a large number of $e^{\pm}$ and finally escape at the photosphere in the form of thermal-like radiation through coherent Compton scattering (CC) and inverse Compton (IC) scattering. 

Under a strong magnetic field condition, photons with different polarization modes (E-mode and O-mode) are produced, and they have different transport approaches and scattering opacity which change with angle \citep{Van2016}. The angle-varying scattering opacity leads to an anisotropic photon escaping intensity. In this case, however, the emergent radiation comes from a small-bubble fireball expanding through a narrow magnetic tube. We thus neglect the angular effect and assume an isotropic relativistic fireball. The corresponding Rosseland mean optical depth for the E-mode photons \citep[][]{TD1995,Meszaros1992herm.book.....M, Lyubarsky2002} of the fireball with a size of $l$ is expressed as

\begin{equation}
 \tau_{\perp} = \frac{4\pi^2}{5}\sigma_{\rm T}\bigg(\frac{kT}{m_ec^2}\frac{B_Q}{B_R}\bigg)^2 n_{\pm}l, \label{eq:E_mode}\\ 
 \end{equation} 

where $\sigma_{\rm T}$ is the Thompson cross section, $m_e$ is the electron mass, c is the light speed, and $B_R$ is the magnetic field at radius R, i.e., $B_R = B_*\big(R/R_s\big)^{-3}$. The O-mode photons generally have a higher optical depth ($\tau_{\parallel} \sim n_{\pm}\sigma_T l$), suggesting that the E-mode photons are easier to escape. The different optical depths of these two polarization states should result in different photosphere radii, which makes it sophisticated to calculate the radiation spectrum. However, there is a detailed balance between these two modes of polarization \citep[][]{Barchas2021}. Due to this, when the E-mode photons escape first, the O-mode photons are likely to be continuously converted into E-mode. On this basis, we simplify our model by adopting a single E-mode photosphere to facilitate the study of the radiation spectrum. By requiring $\tau_{\perp}=1$ in equation (\ref{eq:E_mode}), one can drive the minimal radiation radius of the expanding fireball as $l_x \sim 10^60$ cm, which is about one order of magnitude larger than $l_0$.

\end{enumerate}
	
	\section{Radiation and Fit} \label{sec:RF}
	
Scattering and Absorption Opacities under Strong Magnetic Field

In a strong magnetic field, the Compton scattering and absorption cross-section vary significantly from the case in a non- or weak- magnetic field \citep{Canuto1971PhRvD...3.2303C, Herold1979PhRvD..19.2868H}, giving rise to resonance feature at cyclotron frequency $\omega_B = eB/m_ec$ and its harmonics \citep{Daugherty1981STIN...8130061D}. Within such a framework, the influence of the magnetic field on the scattering cross-section of the two linearly polarized photons should be considered near the magnetar surface. Below $\omega_B$, the scattering cross-section for O-mode photons is close to the classical Thompson value, $\sigma_{\rm T}$, whereas the cross-section for the E-mode photons will be strongly suppressed to $\omega/\omega_B$ times of $\sigma_{\rm T}$, where $\omega = 2\pi/\nu$ is the circular photon frequency. It is worth noting that if both E-mode and O-mode photons propagate parallel to the field lines, their distinction will abate, and they will be represented by the circular polarization mode. In such a case, as long as $\omega<\omega_B$, the scattering cross-sections of photons in all energies will be reduced \citep{Herold1979PhRvD..19.2868H}.

The expression of scattering opacity of linearly polarized photons converted from mode $j$ to mode $i$ ($i, j$=1, 2) in a strong magnetic field can be written as follows \citep{Van2016, HL2003MNRAS.338..233H, Ventura1979PhRvD..19.1684V}
\begin{equation}
 \kappa^{\rm es}_{ji} = \frac{n_{e}\sigma_{\rm T}}{\rho} \sum_{\alpha=-1}^1 \frac{\omega^2}{(\omega+\alpha\omega_B)^2 
 + \Gamma_{e}^2/4} \left|e_\alpha^j \right|^2 A_\alpha^i \quad,
 \label{eq:kappa}
\end{equation}
where $\rho$ is the mass density of the $e^{\pm}$ which an be estimated as $\rho\sim n_em_e$, $\Gamma_e = 2e^2\omega_B^2/(3m_ec^3)$ is the linewidth of the cyclotron resonance. The vector $\boldsymbol{e}^j$ is the normal mode polarization vector, whose cyclic components are 

\begin{align}
 \left| e_{\pm 1}^{\,j} \right|^2 & = \frac{(1 \pm K_j \,{\rm cos}\,\theta_B)^2}{2(1+K_j^2)} \nonumber \\
 \left| e_0^{\,j} \right|^2 & = \frac{K_j^2 \,{\rm sin}^2\,\theta_B}{1+K_j^2}\quad ,
\end{align}
where $\theta_B$ is the angle between the wave propagation direction of the incident photons and the magnetic field vector, $K_j$ is a term encompassing the influences of vacuum and plasma dispersion, whose specific expression and detailed derivation are involved in the work of \citealt{HL2003MNRAS.338..233H}, and $A_\alpha^i$ is the angle integral given by $A_\alpha^i = 3/4 \int \left|e_\alpha^i (\theta_B') \right|^2 {\rm sin}\,\theta_B' \,{\rm d}\theta_B' \quad $ , where the $\theta_B'$ is the angle between the propagation direction of the scattered photon and the magnetic field vector. Thus the electron scattering for a photon in polarization mode j and incident angle $\theta$ to scatter to any polarization mode and angle is given by

\begin{equation}
 \kappa^{\rm es}_{j} \; =\; \frac{n_{e}\sigma_{\rm T}}{\rho} \sum_{\alpha=-1}^1 
 \frac{\omega^2}{(\omega+\alpha\omega_B)^2 + \Gamma_{e}^2/4} \left| e_\alpha^j \right|^2 A_\alpha \quad ,
\label{eq:kappa_esj}
\end{equation}
	
where $A_\alpha = \sum_{i=1}^2 A_\alpha^i$. In the transverse-mode approximation which we employ here, the polarization vector $\boldsymbol{e}^j$ satisfies the completeness relation\footnote{$\sum_{j=1}^2 \left| e_{\pm 1}^{\,j} \right|^2 = (1+\rm{cos}^2\theta_B)/2$ and $ \left| e_0^{\,j} \right|^2 = \rm{sin}^2 \theta_B$}, and thus $A_\alpha = 1$\citep{HL2003MNRAS.338..233H}. Equation (\ref{eq:kappa_esj}) is used to calculate the scattering coefficient in this study, and it greatly simplifies the derivation of radiation spectrum in our model. Under the physical conditions described in Section \ref{sec:physical_picture}, we calculate the analytical expressions of the scattering coefficients ($\kappa^{\rm es}_1$ and $\kappa^{\rm es}_2$, respectively) of E-mode and O-mode photons in equations (\ref{eq:ap1}) and (\ref{eq:ap2}), respectively. Both expressions are parameterized by $B$, $\theta_B$, $kT$, where $B$ is the local magnetic field which can be estimated as $B_r$. We note that $\theta_B$, as the angle between the incident photon and the magnetic field vector, is naturally different for different photons. While $\theta_B$ may have a certain distribution, there seem to be few ways of knowing and verifying that. Therefore, we adopt an effective incident angle $\langle \theta_B \rangle$ to represent the average effect of all photons with the distributed incident angles.

\begin{figure}
 \label{fig:Gaunt}
 \centering
 \includegraphics[width=0.47\textwidth]{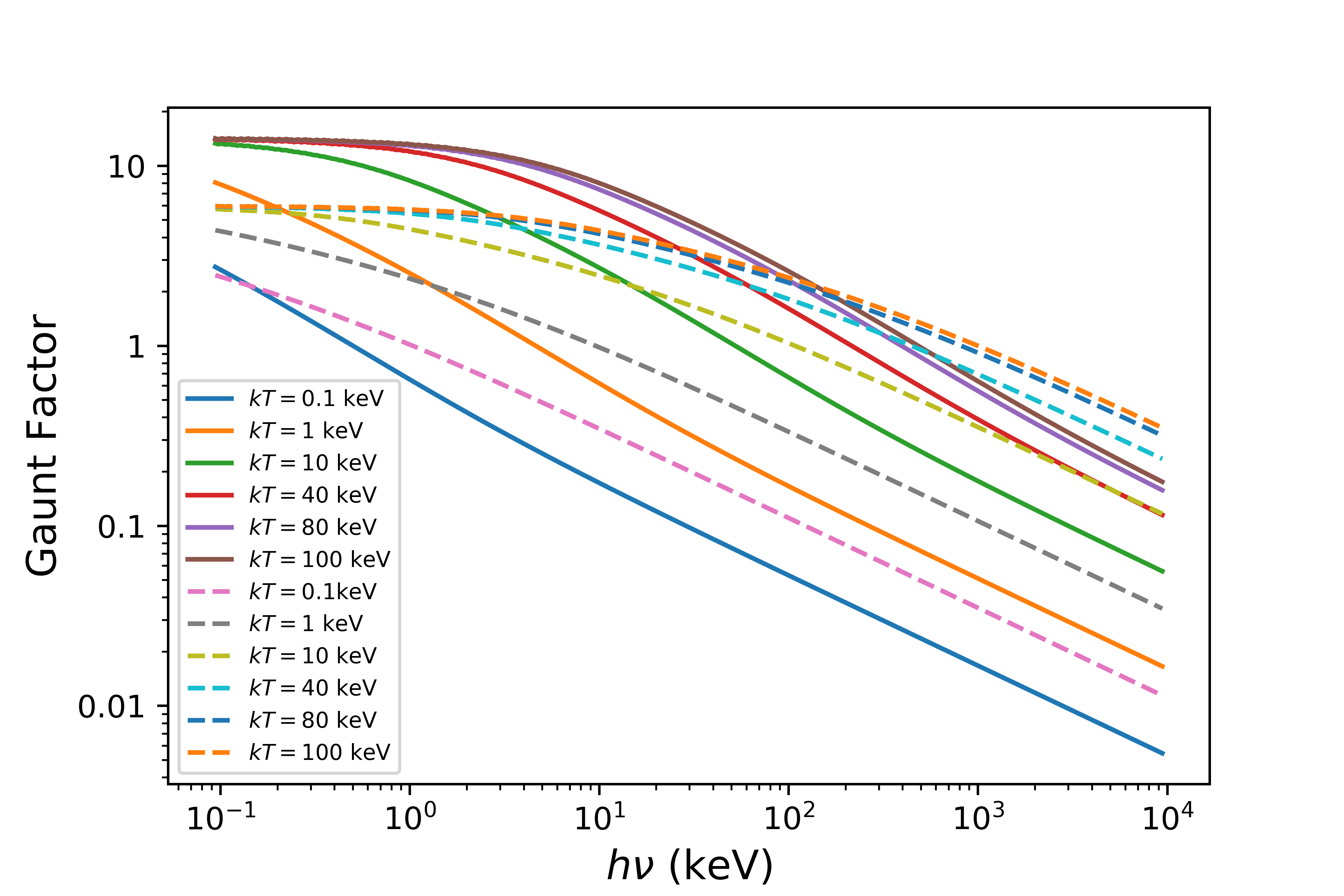}
 \caption{Magnetic Gaunt factors, $g_{\perp}$ (solid) and $g_{\parallel}$ (dashed) as a function of $h\nu$ and $kT$.}
\end{figure}

On the other hand, the electron free-free absorption opacity is given by \citep{Virtamo1975NCimB..26..537V, Pavlov1976JETP...44..300P, Nagel1983A&A...118...66N}

\begin{equation}
 \kappa_{j}^{\mathrm{ff}}=\frac{\alpha_{0}}{\rho} \sum_{\alpha=-1}^{1} \frac{\omega^{2}}{\left(\omega+\alpha \omega_{B e}\right)^{2}+\nu_{\mathrm{e}}^{2}}\left|e_{\alpha}^{j}\right|^{2} \bar{g}_{\alpha}^{\mathrm{ff}}, 
 \label{eq:kappa_ffj}
\end{equation}

where

\begin{equation}
 \begin{aligned} 
 \alpha_{0} &=4 \pi^{2} \alpha_{\mathrm{F}}^{3} \frac{\hbar^{2} c^{2}}{m_e^{2}}\left(\frac{2 m_e}{\pi k T}\right)^{1 / 2} \frac{n_{\pm}^2}{\omega^{3}}\left(1-e^{-\hbar \omega / k T}\right) \\ &=\alpha_{0}^{\mathrm{ff}} \frac{3 \sqrt{3}}{4 \pi} \frac{1}{\bar{g}^{\mathrm{ff}}}
 \end{aligned},
\end{equation}

where $\alpha_{\rm F} = q_e^2\hbar c=1/137 $ is the fine structure constant, and $\alpha_{0}^{\mathrm{ff}}$ and $\bar{g}^{\mathrm{ff}}$ are the free-free absorption coefficient and velocity-averaged free-free Gaunt factor, respectively, in the nonmagnetic environment. In equation (\ref{eq:kappa_ffj}), $\bar{g}_{\pm 1}^{\mathrm{ff}}=\bar{g}_{\perp}^{\mathrm{ff}} $ and $ \bar{g}_{0}^{\mathrm{ff}}=\bar{g}_{\|}^{\mathrm{ff}} $ are the modified free-free Gaunt factors in magnetic field, which can be evaluated by the integral expressions as follows:

\begin{align*}
 \overline{g}_{\perp}(\omega, T)=\int_{-\infty}^{\infty} d x \exp \left(-\frac{\hbar \omega}{k T} \sinh ^{2} x\right) C_{1}\left(\frac{\omega}{\omega_{B}} e^{2 x}\right) 
\end{align*}
\begin{align}
 \overline{g}_{\parallel}(\omega, T)=\int_{-\infty}^{\infty} d x \exp \left(-\frac{\hbar \omega}{k T} \sinh ^{2} x\right) 2 \frac{\omega}{\omega_{c}} e^{2 x} C_{0}\left(\frac{\omega}{\omega_{B}} e^{2x}\right) ,
 \label{eq:Gaunt_M}
\end{align}

where the function $C_0$ and $C_1$ are the Coulomb matrix elements defined by \citealt{Virtamo1975NCimB..26..537V}, and partially solved by \citealt{Ventura1973PhRvA...8.3021V} via some discrete tabulated values. $ C_0$ and $C_1$ can be analytically expressed as

\begin{align*}
 C_{0}(x)=x^{-1}-\exp (x) E_{1}(x) 
\end{align*}
\begin{align}
 C_{1}(x)=(1+x) \exp (a) E_{1}(x)-1 ,
\end{align}

where $ E_{1}(x)=\int_{x}^{\infty} e^t\ d t $ is the exponential integral functions \citep{Abramowitz1988AmJPh..56..958A}. For $\omega \ll \omega_B$ (which is valid in this study), The expression of equation (\ref{eq:Gaunt_M}) can be replaced with \citep{Meszaros1992herm.book.....M}

\begin{align}
 \begin{aligned} 
 g_{\perp}(\omega, T) &=\frac{1}{2} \int_{-\infty}^{\infty} d p \exp \left(-p^{2} / 2 m_e k T\right) \frac{C_{1}\left(a_{+}\right)+C_{1}\left(a_{-}\right)}{\left(p^{2}+2 m_e \hbar \omega\right)^{1 / 2}} \\ 
 g_{\parallel}(\omega, T) &=\int_{-\infty}^{\infty} d p \exp \left(-p^{2} / 2 m_e k T\right) \frac{a_{+} C_{0}\left(a_{+}\right)+a_{-} C_{0}\left(a_{-}\right)}{\left(p^{2}+2 m_e \hbar \omega\right)^{1 / 2}}, \end{aligned} 
 \label{eq:Gaunt_simp}
\end{align}

where $ a_{\pm}=\left(p \pm\left[p^{2}+2 m_e \hbar \omega\right]^{1 / 2}\right)^{2}(2 m_e \hbar \omega)^{-1} $. equation (\ref{eq:Gaunt_simp}) appears to be unsolvable with an analytical solution. We thus obtain a wide range of tabulated values by numerical calculation. Fig \ref{fig:Gaunt} shows the evolution of this dimensionless magnetized Gaunt factor in our applicable range of $h\nu$ and $kT$.

\subsection{Radiation Transfer Equation and Flux Calculation}
Our model predicts that the thermal photons of expanding fireball are subject to the coherent CC and incoherent IC scattering, so the resulted spectrum is likely a modified blackbody (hereafter, MB\footnote{We use ``MB" to distinguish from multicolor blackbody \citep[][]{Meng2018,Meng2019,Meng2021,Wang2021,Meng2022}.}) shape with an IC tail at the high-energy end. A parameterized function of the model-predicted spectrum can be derived by considering the above two radiation processes.

Whether the CC or IC dominates the radiation process is determined by the Compton $y$ parameter, which varies at different photon energies. Following \cite{Rybicki1986}, we define \begin{equation}
y \equiv \frac{kT \overline{N}}{m_e c^2},
\end{equation}
where $k$ is the Boltzmann constant, and $\overline{N}$ is the mean scattering number when a photon travels a free path.

Depending on the $y$ parameter, the derived flux can be considered in the following two regimes:

\begin{itemize}
 \item For $y<$1, CC dominates, and the energy of a scattered photon is not significantly changed. In this case, an MB spectral shape is expected. The specific intensity of CC is closely related to the scattering and absorption of the thermal photons \citep[][]{Rybicki1986}:
\begin{equation}
I^{\rm CC}_{\nu} = \frac{2B_{\nu}}{1+\sqrt{(\kappa_{\rm ff}+\kappa_{\rm es})\kappa_{\rm ff}^{-1}}}, \label{MBB}
\end{equation}
where $\kappa_{\rm ff} = \kappa^{\rm ff}_1 + \kappa^{\rm ff}_2$ is the linear superposition of absorption factor for E-mode and O-mode photons of the free-free (bremsstrahlung) process in a thermal medium, $\kappa_{\rm es} = \kappa^{\rm es}_1 + \kappa^{\rm es}_2$ is the superimposed scattering opacity from $e^{\pm}$ plasma, and $B_\nu$ is the Planck function expressed as
\begin{equation}
B_{\nu} = \frac{2h\nu^3/c^2}{\mathrm{exp}(h\nu/kT)-1}, \label{BB}
\end{equation}
where $h$ is the Planck constant.

Considering the Doppler boosting, the observed specific flux at a luminosity distance, $D_{\rm L}$, can be calculated by

\begin{align}
F^{\rm CC}_{\nu_{\rm obs}} &= \frac{4\pi D^3}{c^2}\frac{h(\nu_{\rm obs}/D)^3}{\mathrm{exp}\bigg(\frac{h\nu_{ obs}}{DkT}\bigg)-1}\Bigg(1+\sqrt{\frac{\kappa_{\rm ff}+\kappa_{\rm es}}{\kappa_{\rm ff}}}\Bigg)^{-1} \nonumber\\ 
&\quad \times \bigg(\frac{l_x}{D_{\rm L}}\bigg)^2, \label{eq:MBF}
\end{align}
where $D$ is the Doppler factor. 
$F_{\nu_{\rm obs}}$ strongly depends on the competition between $\kappa_{\rm ff} $ and $\kappa_{\rm es}$, which can be easily noticed by introducing a characteristic frequency, $\nu_{\rm 0, CC}$, at which the scattering and absorption opacity are equivalent, i.e.
\begin{equation}
\kappa_{\rm es} = \kappa_{\rm ff}(\nu_{\rm 0, CC}). \label{es_ff_x0}
\end{equation}

For $\nu < \nu_{\rm 0, CC}$, absorption is more dominant, and equation (\ref{MBB}) approaches the Rayleigh-Jeans limit and the observed spectrum [equation (\ref{eq:MBF})] is a pure blackbody. On the other hand, for $\nu > \nu_{\rm 0, CC}$, the scattering gradually becomes significant (see Fig. \ref{fig:kff_kes} left) at higher energies and modifies the spectrum (equation \ref{eq:MBF}) significantly.

\begin{figure*}
 \centering
 \includegraphics[width=1.0\textwidth]{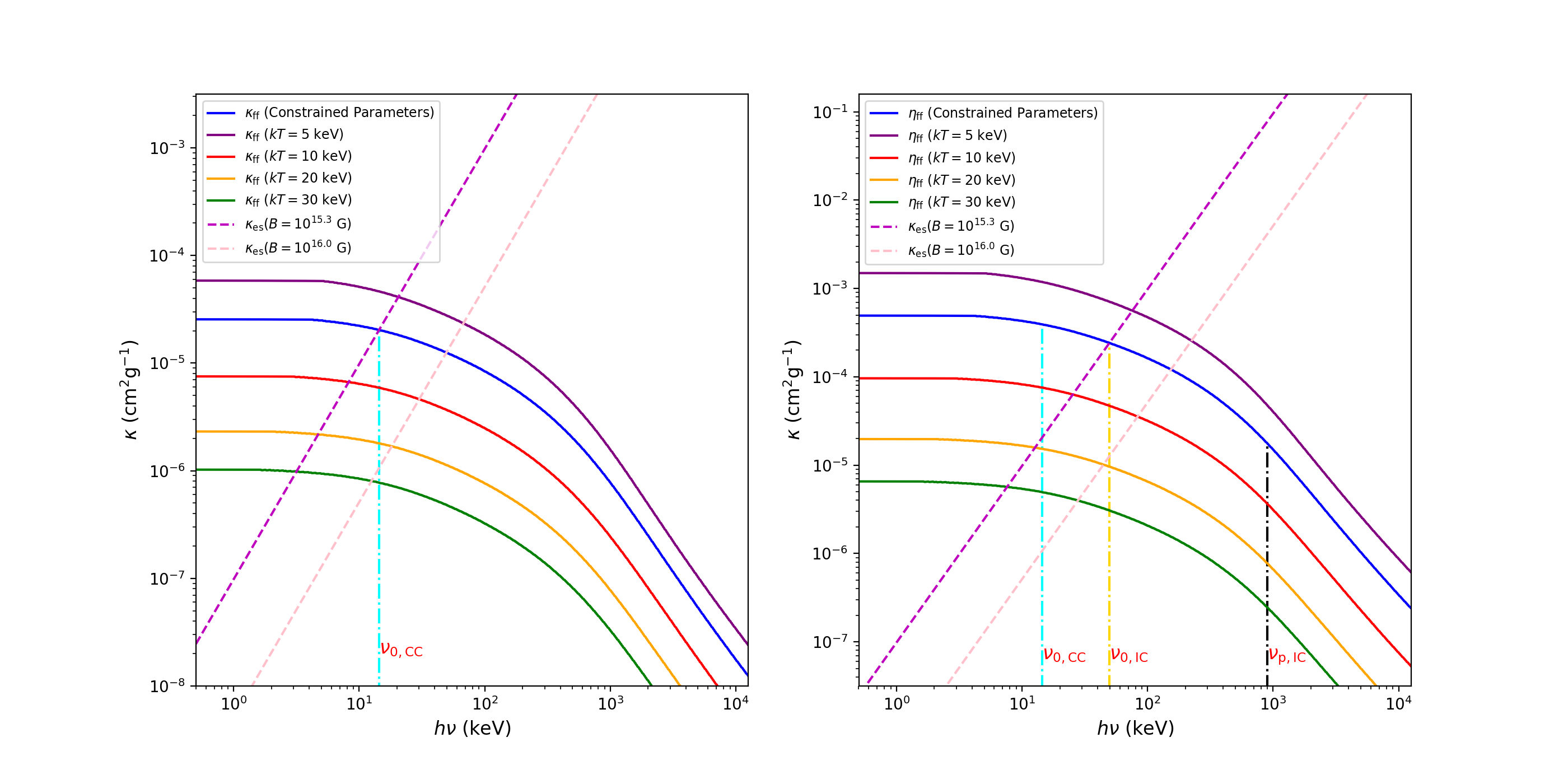}
 \caption{Absorption and scattering coefficients characterized with the Coherent Compton (CC) and Inverse Compton (IC) frequencies. \textit{Left}: absorption opacity, $\kappa_{\rm ff}$, as a function of photon energy at different equilibrium temperatures. $\kappa_{\rm ff}$ evolves with the temperature, marked by the solid lines of different colors. The blue line is calculated using the constrained parameters of integral spectrum (see Table \ref{tab:MB_Wien} ). The dashed line represents the $\kappa_{\rm es}$, the values of which vary with the strength of the magnetic field (marked with different colors). The cyan vertical line (dotted-dashed) marks the $\nu_{\rm 0, CC}$ at $kT\sim6.63$ keV. \textit{Right}: the solid line represents the curves following $\eta_{\rm ff} = (\frac{m_e c^2}{4kT})\kappa_{\rm ff}(\nu_{\rm 0, IC})$. $\nu_{\rm 0, IC}$ is marked with gold vertical line (dotted-dashed), and $\nu_{\rm p, IC}$ is marked with black vertical line (dotted-dashed). The lines of $\kappa_{\rm es}$ and $\nu_{\rm 0, CC}$ are the same as those in the \textit{Left} panel.}
 \label{fig:kff_kes}
\end{figure*}

As shown in equations (\ref{eq:ap1}) and (\ref{eq:ap2}), $\kappa_{\rm es,1}$ and $\kappa_{\rm es,2}$ are dependent on four parameters, namely, $n_{\pm}$, $B$, $\theta_B$ and $\nu$. Our calculation shows that $B$ and $\nu$ are dominant in those equations. This can be seen in Fig. \ref{fig:kff_kes}, which shows the plot of $\kappa_{\rm es}$ as a function of $B$ and $\nu$. The plot shows that for a pair plasma with a density of $n_{\pm} = 2.4\times 10^{23}$ cm$^{-3}$ and temperature in the range of 10-60 keV under a magnetic field of $3\times 10^{16}$ G, it is calculated that $h\nu_{\rm 0,CC}$ is typically $\gtrsim$17 keV in the co-moving frame. Such a frequency falls well into the MeV range in the observer frame, confirming that the CC process is indeed necessary to be taken into account. This can also be verified by checking $x_{\rm 0, CC} \equiv h\nu_{\rm 0, CC}/kT\ll 1 $, suggesting that a photon at such a characteristic frequency is significantly subject to scattering in a hot plasma.

\item For $y$ $\gg$ 1, the radiation is dominated by IC, and the energy is transferred from $e^{\pm}$ pairs to the photons. In this case, the specific intensity described by equation (\ref{eq:MBF}) will naturally transition to the Wien law \citep[][]{Rybicki1986}:

\begin{equation}
I^{\rm IC}_{\nu} = \frac{2h\nu^3}{c^2}e^{-\alpha}e^{-x}, \label{eq:Wien}
\end{equation}
where the factor $e^{-\alpha}$ is constant related to the local rate at which photons are produced, defined by \cite[][]{Landau1969}
\begin{equation} 
e^{-\alpha} = \frac{N}{V}\bigg(\frac{h^2}{2\pi mkT}\bigg)^{3/2} \sim n_{c}\bigg(\frac{h^2}{2\pi m_{c}kT}\bigg)^{3/2}, \label{alpha}
\end{equation}
where m is the particle mass, N is the total number of particles, and V is the volume of the radiation region. $n_c$ and $m_c$ are defined as coupling number density and coupling mass, respectively. $e^{-\alpha}$ measures degeneracy of the particles. For example, $e^{-\alpha}$ is negligible if the particles meet the classical limit condition and are entirely non-degenerate. Otherwise, $e^{-\alpha}$ can be a significant number if the photons are coupling with the $e^{\pm}$ pairs in a degenerate plasma. The specific flux at the observer is
\begin{equation}
F^{\rm IC}_{\nu_{\rm obs}} = 
\frac{2\pi D^3 h\nu_{\rm obs}^3}{{c^2}}e^{-\alpha}\mathrm{exp}\bigg(\frac{-h\nu_{\rm obs}}{DkT}\bigg)\bigg(\frac{l_x}{D_\mathrm{L}}\bigg)^2. \label{eq:ICF}
\end{equation}
Another critical frequency, $\nu_{\rm 0,IC}$, is defined by requiring $y(\nu_{\rm 0,IC}) = 1$, above which IC process becomes significant. For pairs which follow the non-relativistic\footnote{In the comoving frame, the mean Lorentz factor of the $e^{\pm}$ pairs, $\left \langle \gamma_{\pm} \right \rangle$, can be estimated by $\left \langle \gamma_{\pm} \right \rangle = \sqrt{(12(kT/m_e c^2)^2)}$. For a characteristic temperature in our model, $T_c \sim 15$ keV, $\left \langle \gamma_{\pm} \right \rangle \ll 1$.} thermal distributions in the comoving frame, $\nu_{\rm 0,IC}$ satisfies the relationship \citep[][]{Rybicki1986}:
\begin{equation}
\kappa_{\rm es} = \bigg(\frac{m_ec^2}{4kT}\bigg)\kappa_{\rm ff}(\nu_{\rm 0, IC}). \label{es_ff_coh}
\end{equation}
The IC process will dominate the spectrum at a photon energy around $h\nu_{\mathrm{p, IC}} = 3kT$, which forms the high-energy Wien tail (Fig. \ref{fig:kff_kes}). 
\end{itemize}
The final observed flux can be written as the following function:
\begin{align}
F_{\nu_{\rm obs}} = F^{\rm CC}_{\nu_{\rm obs}}(n_{\pm}, kT, B_{*}, l_0, \langle \theta_B \rangle) + F^{\rm IC}_{\nu_{\rm obs}} (kT,\alpha,l_0), \label{eq:total_flux}
\end{align}
which can be used to directly fit to the observed data. Note that we assumed an on-axis observer for equation (\ref{eq:total_flux}) so the Dopper factor, $D=1/[\Gamma(1 - \beta \mathrm{cos} \theta)] \sim \Gamma$, is replaced with the bulk Lorentz factor $\Gamma$ of the pair plasma in our calculations. In the fitting process, $\Gamma$ can be calculated by substituting the photosphere radius obtained by equation (\ref{eq:E_mode}) into equation (\ref{eq:Gamma_T}).

Considering the physical conditions, the priors and allowed ranges of the five free parameters in equation (\ref{eq:total_flux}) are set up as follows:

\begin{itemize}
 \item $n_{\pm}$: The number density of the $e^{\pm}$ in the emission region. Log-uniformed distributed in range [$10^{20.0}, 10^{28.0}$].
 \item $T$: The thermodynamic equilibrium temperature in the co-moving reference. $kT$ is uniformed distributed in range [0.01, 100.0].
 \item $B_{*}$: The local surface magnetic field of the magnetar. Log-uniformed distributed in range [$10^{14.0}, 10^{17.0}$]
 \item $l_0$: The initial radius of the expanding fireball. Log-uniformed distributed in range [$10^{3.0}, 10^{6.0}$].
 \item $\langle \theta_B \rangle$: The average incident angle between the photons and magnetic field vector before scattering. Uniformed distributed in range [0.0, $\pi/2$].
\item $\alpha$: The index related to the IC intensity. Uniformed distributed in range [0.0, 10.0] so $e^{-\alpha}$ is a significant number in range of [$4\times10^{-5}$, 1].
 
\end{itemize}

\begin{figure*} 
\label{fig:fitting}
\centering
\subfigure{
\begin{minipage}[t]{0.3\linewidth}
\centering
\includegraphics[width=5.5cm]{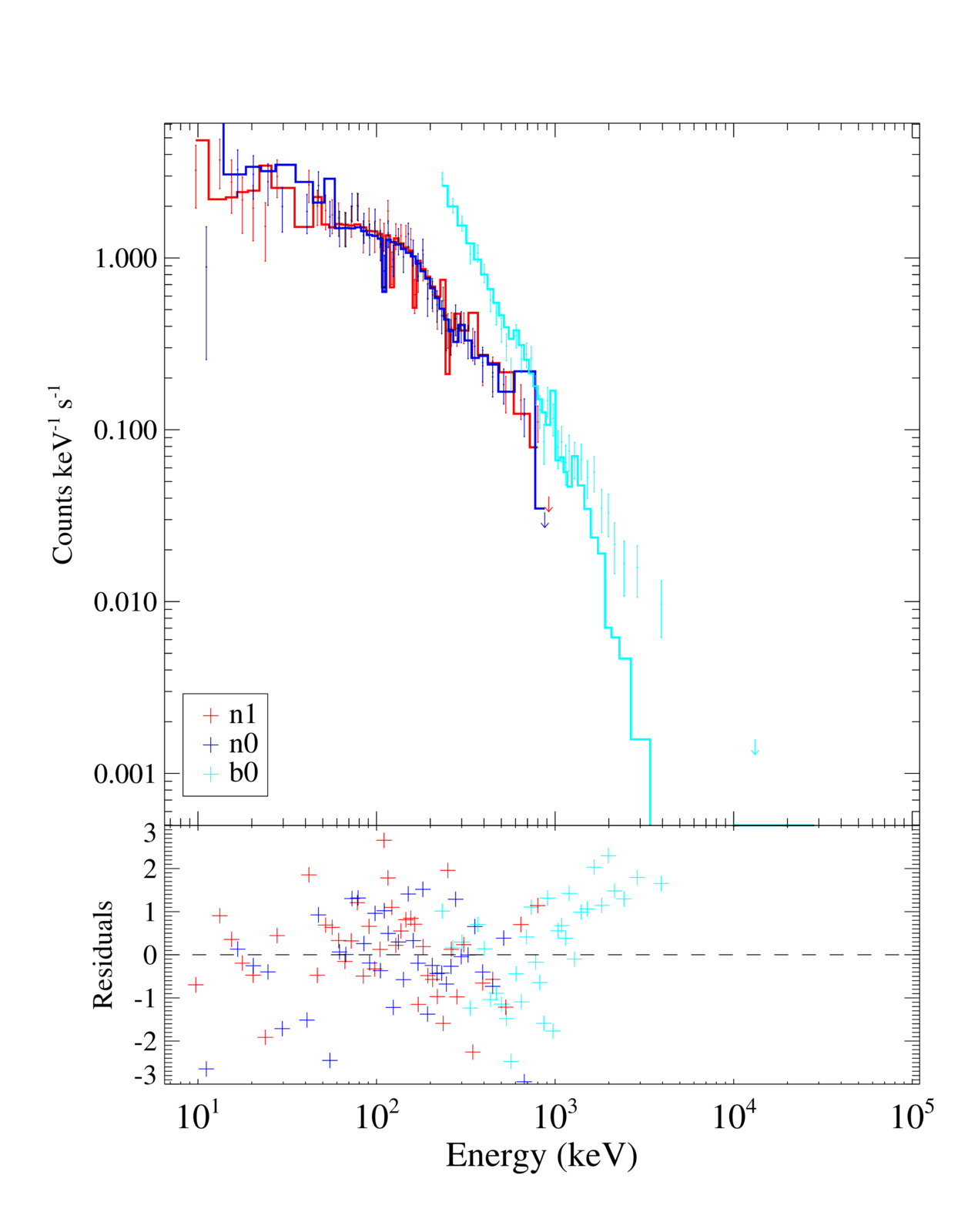}
\hspace{.2in} 
\end{minipage}%
}%
\subfigure{
\begin{minipage}[t]{0.3\linewidth}
\centering
\includegraphics[width=5.5cm]{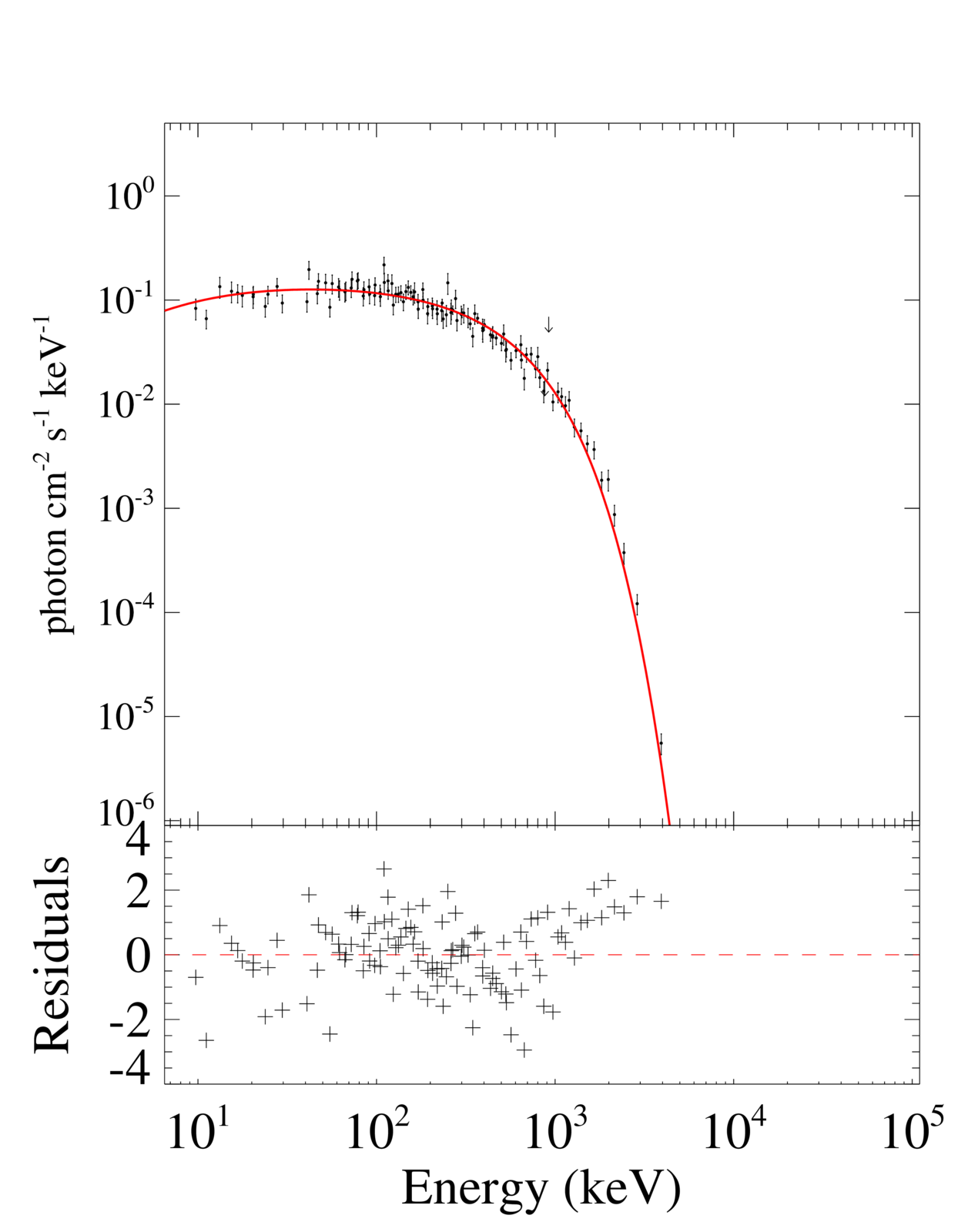}
\hspace{.2in}
\end{minipage}
}
\subfigure{
\begin{minipage}[t]{0.3\linewidth}
\centering
\includegraphics[width=5.5cm]{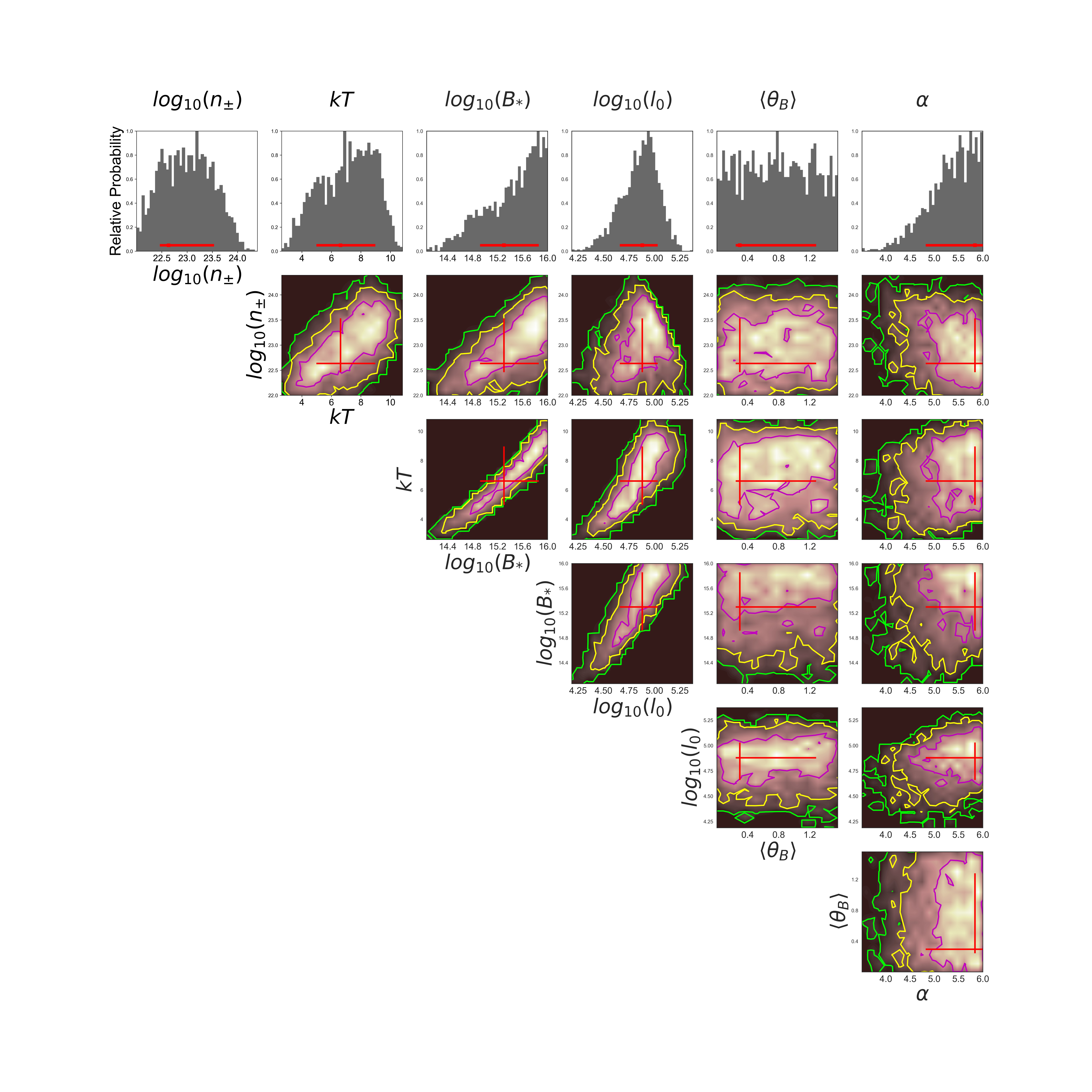}
\hspace{.2in}
\end{minipage}%
}
\centering
\caption{Time-integrated spectral fit of our model to the observed spectra of GRB 200415A between $T_0 - 0.005 s$ and $T_0 + 0.20 s$. {\it Left}: observed photon count spectra over-plotted with best-fit model; {\it Middle}: Deconvolved model-predicated photon spectrum; {\it Right}: corner diagram of the parameter constraints. Histograms show the 1-D likelihood distribution of the fitting parameters, and contours show the 2-D likelihood map constrained by the MCMC method. Red crosses mark the best-fit values, and contours represent the 1-,2-,3-$\sigma$ regions. } 
\end{figure*}

\begin{table*}
		\begin{center}
			\caption{The spectral fitting results of GRB 200415A}
			\label{tab:MB_Wien}
			 \setlength\tabcolsep{1.9pt}
 \begin{tabular}{cc|c|ccccccc}
				\hline
				\hline
 	Time Interval (s)& &Flux & & &Model Parameters & &\\
				
				\cline{1-2} \cline{4-10}
				
				$t_1$&$t_2$&(erg cm$^{-2}$s$^{-1}$)&$log_{10}(n_{\pm}/$cm$^{-3})$&$kT$ (keV)&$\mathrm{log}_{10}(B_{*})/\mathrm{G}$& $log_{10}(l_0/\mathrm{cm}$)& $\langle \theta_{B} \rangle$&$\alpha$&PGSTAT/dof\\
				\hline

				-0.005& 0.20&$3.90_{-0.31}^{+0.27}\times10^{-5}$&$22.63_{-0.17}^{+0.90}$ & $6.62_{-1.63}^{+2.36}$&$15.30_{-0.38}^{+0.56}$ &$4.88_{-0.22}^{+0.15}$&$0.30_{-0.05}^{+0.98}$&$5.84_{-1.02}^{+0.16}$&277.5/349\\
				\hline
				-0.005& -0.001&$2.90_{-0.39}^{+0.44}\times10^{-4}$&$22.31_{-0.20}^{+0.45}$ & $2.48_{-0.69}^{+0.15}$&$15.83_{-0.51}^{+0.08}$ &$5.21_{-0.23}^{+0.01}$&$1.57_{-1.29}^{+0.01}$&$4.96_{-1.15}^{+0.65}$&262.2/349 \\
				-0.001& 0.01&$2.36_{-0.34}^{+0.38}\times10^{-4}$&$22.60_{-0.20}^{+1.16}$ & $9.64_{-3.84}^{+0.52}$&$15.79_{-0.80}^{+0.08}$ &$5.04_{-0.53}^{+0.18}$&$1.39_{-1.14}^{+0.18}$&$5.71_{-1.43}^{+0.02}$&276.1/349 \\
				0.01& 0.04&$7.99_{-0.97}^{+1.12}\times10^{-5}$&$24.34_{-0.74}^{+0.44}$ & $4.55_{-1.05}^{+2.79}$&$15.28_{-0.52}^{+0.51}$ &$4.35_{-0.21}^{+0.23}$&$0.36_{-0.06}^{+0.96}$&$2.27_{-1.12}^{+3.03}$&249.3/349\\
				0.04& 0.12&$2.42_{-0.32}^{+0.30}\times10^{-5}$&$24.15_{-0.31}^{+0.31}$ & $5.24_{-2.20}^{+1.50}$&$15.45_{-0.81}^{+0.38}$ &$4.60_{-0.23}^{+0.24}$&$1.16_{-0.90}^{+0.11}$&$4.34_{-1.36}^{+1.17}$&238.5/349 \\
				0.12& 0.20&$1.05_{-0.50}^{+0.60}\times10^{-8}$&$22.09_{-0.09}^{+2.48}$ & $5.26_{-3.63}^{+6.22}$&$15.94_{-1.03}^{+0.06}$ &$5.80_{-1.14}^{+0.02}$&1.14$_{-0.86}^{+0.15}$&$0.33_{-0.00}^{+4.21}$&172.0/349\\
			 \hline
				\hline
		 \end{tabular}
		\end{center}
\end{table*}

We then collected the spectral data of GRB 200415A and performed the spectral analysis on the main burst region between 0.005 and 0.20 s, following the procedure of \cite{YJ2020}. The data reduction follows the standard procedure described in \citep{ZBB2011,ZBB2016,ZBB2018NatAs}. Both time-integrated and time-dependent spectral analyses have been performed, between $T_0 - 0.005 s$ and $T_0 + 0.20 s$. The spectral fitting slices are presented in Table \ref{tab:MB_Wien}. By employing the self-developed spectral fitting package, MCSPECFIT \citep{ZBB2018NatAs}, we were able to fit the spectral data using our model in equation (\ref{eq:total_flux}). The time-integrated fit fit is shown in Fig. \ref{fig:fitting}. The best-fit parameters, along with the corresponding energy flux in each slice, are listed in Table \ref{tab:MB_Wien}. 

The values of the goodness of fit (PGSTAT/dof) in Table \ref{tab:MB_Wien} indicate that our model successfully explains the observation. The values of best-fit parameters are overall consistent with the theoretical predictions as highlighted below: 
\begin{itemize}
 \item $n_{\pm}$ is constrained to a relatively stable value between $10^{23}$ - $10^{24.3}$ cm$^{-3}$, which is well consistent with the estimation in equation (\ref{eq:num}).

 \item $kT$ shows no significant variance in each time interval with a peak value at $kT= 9.64_{-0.52}^{+0.59}$ keV.
 
 \item $l_0$ is stably constrained at a range of $\sim 2$-$10\times 10^4$ cm, which is consistent with model estimation in equation (\ref{eq:l0}).
 
 \item $B_{*}$ is constrained within the range of $\sim$ [15.30, 15.94], which is well consistent with our physical predictions in Section \ref{sec:physical_picture}.
 
\item $\langle \theta_{B} \rangle$ is not well constrained in the range of [0, $\pi/2$], which indicates that the incident angle of the scattering photons relative to the magnetic field vector has little influence on the photon flux and spectral shape. An alternative physical possibility is that due to the asymmetry and strong distortion of the magnetic field, $\theta_{B}$ has an approximately random distribution.
\end{itemize}

\begin{figure*}
 \label{fig:ep_eiso}
 \centering
 \includegraphics[width=0.55\textwidth]{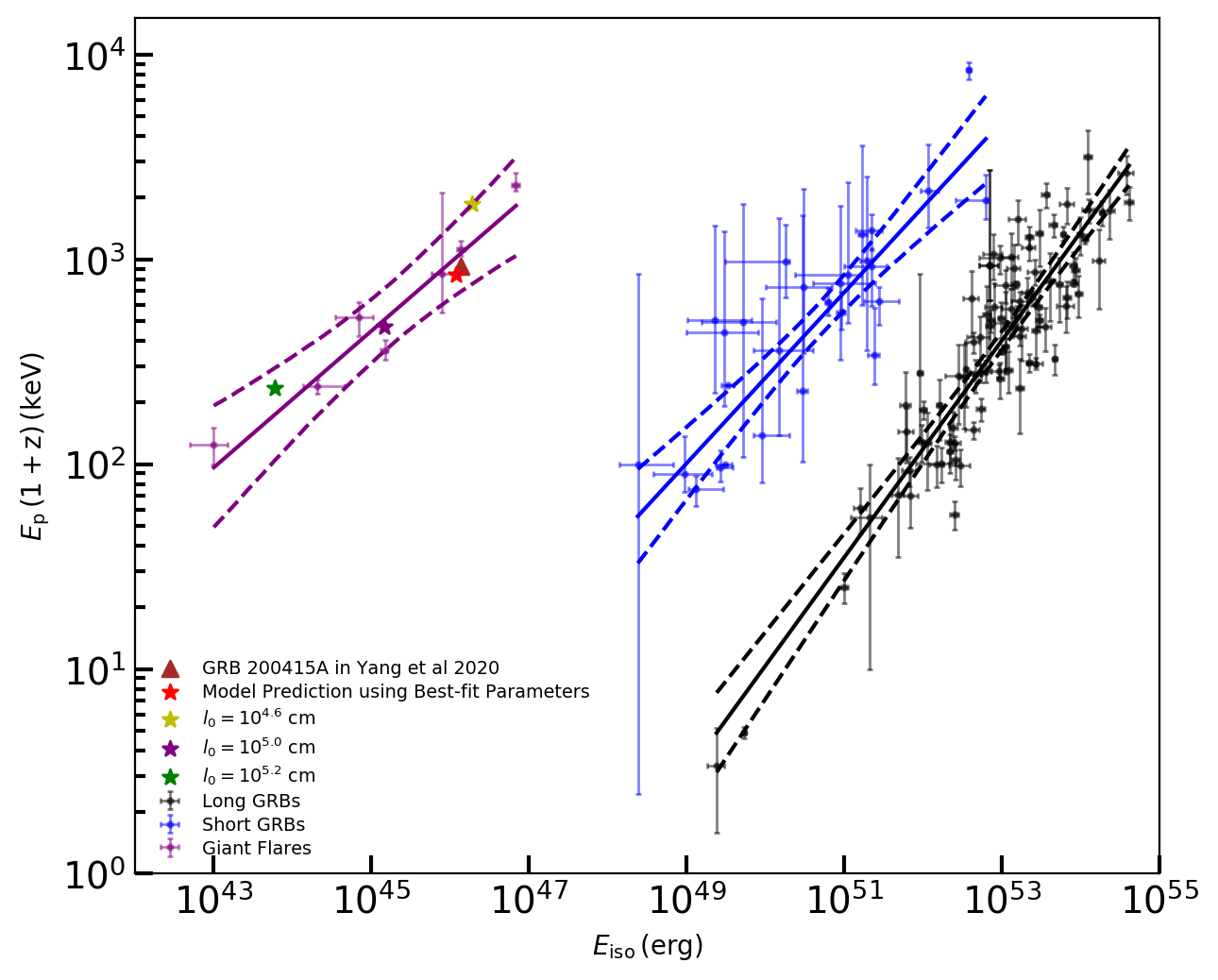}
 \caption{GRB 200415A in $E_{\rm p}$-$E_{\rm iso}$ correlation. The blue, black and purple solid lines represent the best-fit correlations for short, long and MGF populations, respectively. The brown triangle represents the position of GRB 200415A in \citet{YJ2020}. The red star marks the model position for the constrained parameters ($l_0\sim 10^{4.88}$ cm) of the integrated spectrum. Theoretical points with different initial fireball radii $l_0$, and fixed other parameters as constrained, are shown by the stars in other colors.}
\end{figure*}

Our results can also be used to constrain the bulk Lorentz factor, which is $\Gamma \sim (l_x/l_{0})^{3/2} \sim 45.1 $. Such a result is consistent with the lower limit (i.e., $\Gamma>6$) given by \cite{RobertNA2021}. Moreover, the observed temperature can be estimated by $kT_{\rm obs} = \Gamma kT \sim 298.6^{+106.3}_{-73.5}$ keV, which is compatible with the initial temperature $kT_{\rm ini} \sim 328.6^{+18.36}_{-18.56}$ keV obtained in \citet{YJ2020}. Such a high temperature naturally expects a high peak energy \citep[$E_{\rm p}\sim2.82 kT_{\rm obs} = 841.95^{+300.1}_{-207.3}$ keV; ][]{ZhangB2012a}, which is also in good agreement with the peak energy of 
$E_{\rm p} = 926.68^{+51.78}_{-52.33}$ keV from \cite{YJ2020}\footnote{Using a different spectral fitting tool, \cite{ZhangHM2020} obtained a slightly different yet consistent value of $E_{\rm p} = 916.3^{+124.5}_{-112.8} $ keV.}. The isotropic energy predicted by our model can be derived as $E_{\rm iso} = 4\pi D_L^{2}F_{\rm obs}\Delta t/(1+z) \sim 1.17^{+0.08}_{-0.09}\times10^{46}\mathrm{\ erg}$, where $F_{\rm obs}$ is the model flux calculated using the best-fit parameters listed in  Table \ref{tab:MB_Wien}), $\Delta t$ is the time interval of the time-integrated spectrum. Those $E_{\rm p}$ and $E_{\rm iso}$ values (plotted as a red star in Fig. \ref{fig:ep_eiso}) are well consistent with those presented in \cite{YJ2020} and follows the MGF track in the $E_{\rm p}$-$E_{\rm iso}$ diagram \citep[a.k.a. Amati relation;][]{Amati2002}. In addition, our model can reproduce the other MGF-GRBs by varying input parameters such as $l_0$ (Fig. \ref{fig:ep_eiso}), which implies a common origin for the sample.

\begin{figure*}
 \label{fig:kff_fitting}
 \centering
 \includegraphics[width=1.0\textwidth]{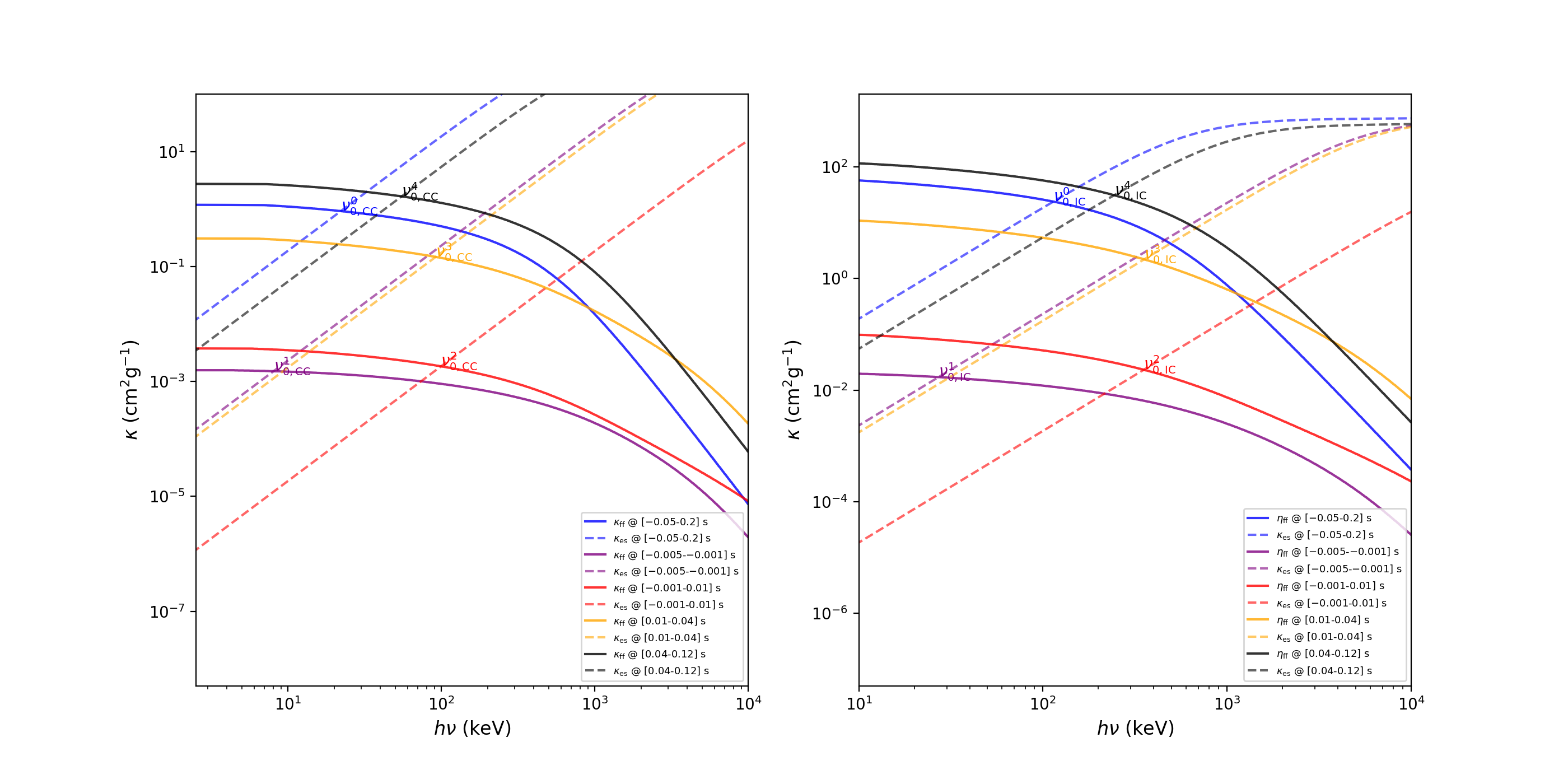}
 \caption{The curves of $\kappa_{\rm ff}$, $\kappa_{\rm es}$ and $\eta_{\rm ff}$ at each time interval between $T_0-0.05$ s and $T_0+0.12$ s, which are obtained from different constrained parameters in Table \ref{tab:MB_Wien}. \textit{Left}: $\kappa_{\rm ff}$ and $\kappa_{\rm es}$. \textit{Right}: $\eta_{\rm ff}$ and $\kappa_{\rm es}$ }
\end{figure*}

In Fig. \ref{fig:kff_fitting}, we plot the $\kappa_{\rm ff}$, $\kappa_{\rm es}$ and $\eta_{\rm ff}$ curves using best-fit values of the time-resolved spectra for each time interval listed in Table \ref{tab:MB_Wien}. We then obtain $h \nu^i_{\rm 0 ,CC}\sim$[22.08, 8.04, 99.22, 91.50, 54.97] keV, $h \nu^i_{\rm 0 ,IC}\sim$[113.86, 27.09, 352.41, 354.01, 243.92] keV, $h \nu^i_{\rm p ,IC}\sim 3\cdot kT^i_{\rm obs}\sim$[549.09, 1313.01, 829.95, 678.40, 654.38] keV, corresponding to the time intervals of [$-0.005$-$0.20$, $-0.005$-$-0.001$, $-0.001$-$0.01$, $0.01$-$0.04$, $0.04$-$0.12$] s. Such values of $h \nu^i_{\rm 0 ,CC}$ and $h \nu^i_{\rm p ,IC}$ confirm that the MB component caused by CC scattering are dominant in each spectrum observed in the $\sim$ MeV range.

\begin{figure}
 \label{fig:SC}
 \centering \includegraphics[width=0.47\textwidth]{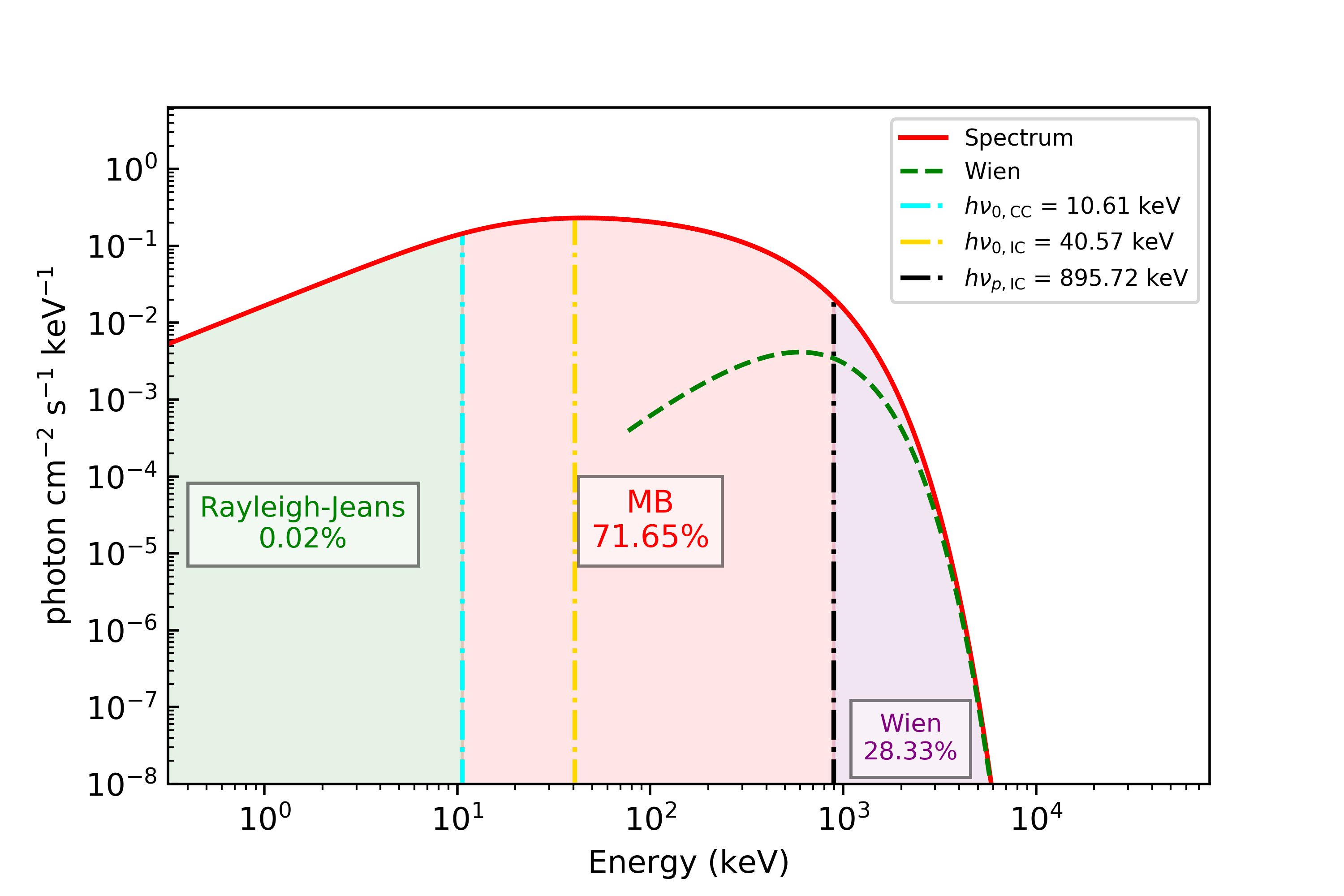}
 \caption{Schematic diagram for energy fractions of different components, within the best-fit time-integrated model spectrum. The green shadow ($\mathcal{F}_{\rm RJ} = 0.02\%$) represents the Rayleigh-Jeans dominant regime. The magenta shadow ($\mathcal{F}_{\rm MB} = 71.65\%$) represents the MB region, dominated by coherent Compton scattering. And the purple shadow ($\mathcal{F}_{\rm Wien} = 28.33\%$) is dominated by the Wien spectrum.}
\end{figure}

\begin{figure*}
 \label{fig:ne_F}
 \centering
\includegraphics[width=1.0\textwidth]{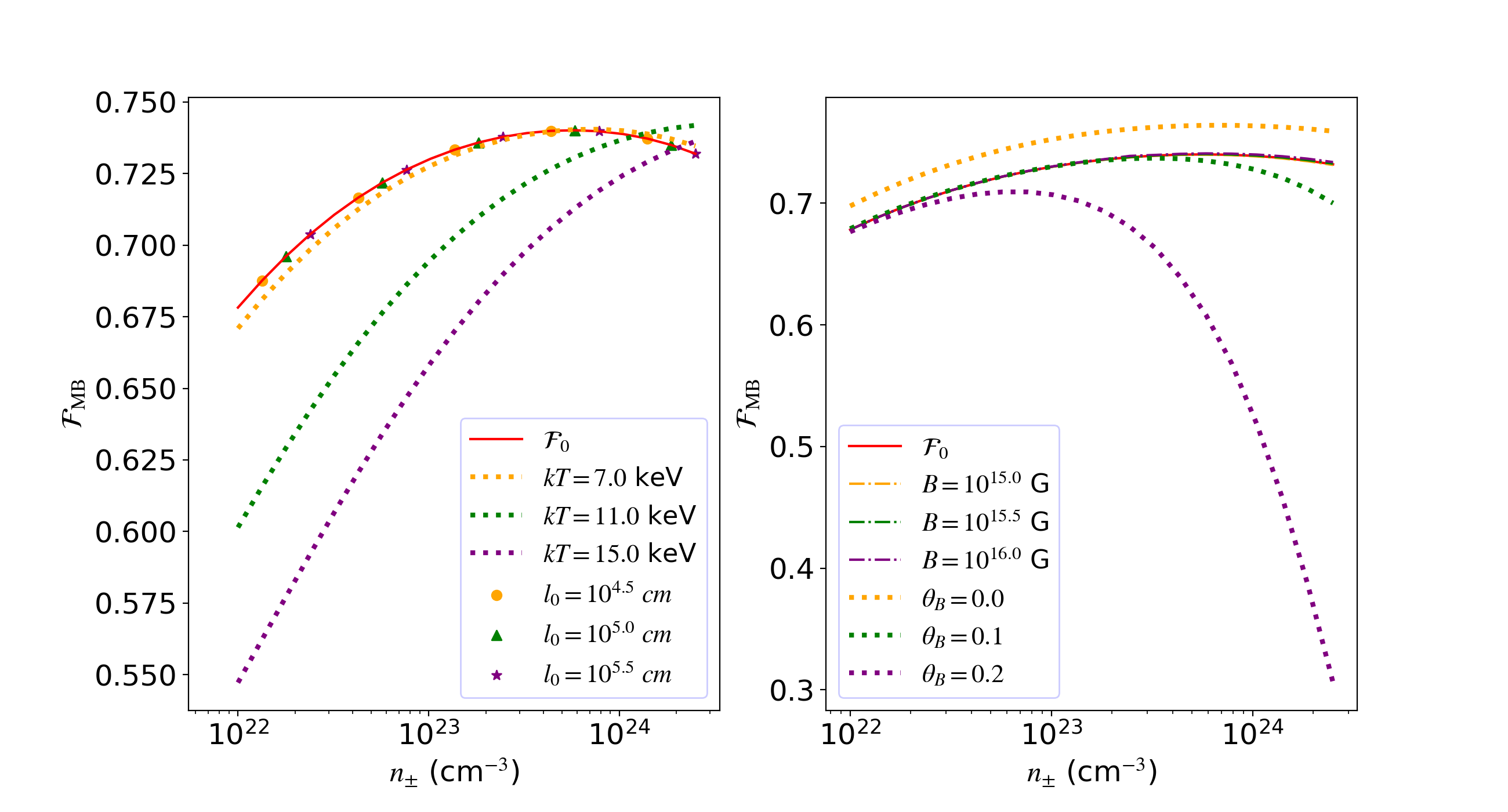}
 \caption{The dependence of the MB component fraction ($\mathcal{F}_{\rm MB}$) on $n_{\pm}$, $l_0$, $kT$, $B_*$ and $\theta_B$. The red solid lines ($\mathcal{F}_0$) are plotted using the best-fit parameters (except for $n_{\pm}$, as a variable) of the time-integrated spectrum in Table \ref{tab:MB_Wien} ($l_0 = 10^{4.88}$ cm, $kT = 6.62$ keV, $B_* = 10^{15.30}$ G, $\theta_B = 0.30$ and $\alpha = 5.84$). \textit{Left}: $\mathcal{F}_{\rm MB}$ dependence on $n_{\pm}$ (solid line), $l_0$ (markers), and $kT$ (dashed lines). \textit{Right}: $\mathcal{F}_{\rm MB}$ dependence on $B$ (dot-dashed lines) and $\theta_B$ (dashed lines).}
\end{figure*}

Fig. \ref{fig:SC} illustrates the spectral components that dominate different regions in the time-integrated spectrum of the best-fit model. Three regions can be identified based on the dominant component of the spectrum: (1) a low-energy region ($\nu \lesssim \nu_{0, \rm CC}$) characterized by the Rayleigh-Jeans component with an energy fraction of $\mathcal{F}_{\rm RJ}$; (2) an intermediate-energy region (between $\nu_{0, \rm CC}$ and $\nu_{p, \rm IC}$) dominated by the MB with an energy fraction of $\mathcal{F}_{\rm MB}$; (3) a high-energy tail ($\nu \gtrsim \nu_{\rm p, IC}$) dominated by Wien's law with an energy fraction of $\mathcal{F}_{\rm Wien}$. By integrating the flux over corresponding energy ranges in these regions with our parameterized model function, we can calculate the energy fractions of each region, which yields $\mathcal{F}_{\rm RJ} = 0.02\%$, $\mathcal{F}_{\rm MB} = 71.65\%$, and $\mathcal{F}_{\rm Wien} = 28.33\%$. We note that $\mathcal{F}_{\rm MB}$ is of our particular interest since it reveals how much the pure blackbody spectrum is modified by coherent and incoherent scattering. The parameter dependence of $\mathcal{F}_{\rm MB}$ is illustrated in Fig. \ref{fig:ne_F}, which shows that the MB component contributes more when the temperature drops, or the incident angle decreases. On the other hand, $\mathcal{F}_{\rm MB}$ weakly depends on the magnetic field and the initial size of the fireball bubble.
\section{Discussion} \label{sec:discussion}
\subsection{Baryon Loading}
As we demonstrate in the following, significant contamination of baryons in the magnetosphere would result in an increase of the photosphere radius, which would result in a longer burst duration. Besides, the peak energy $E_{\rm p}$ should be significantly reduced due to the adiabatic expansion. Thus, the observed short time-scale ($\sim6$ ms) and high $E_{\rm p}$ of the spike emission \citep{YJ2020} point toward a low baryon loading. 
At a certain radius, the Thompson scattering optical depth due to baryon-related electrons can be calculated by \citep{2006ApJ...642..995P}
 \begin{equation}
    \tau_{b} = \frac{1}{\Gamma}r_in_e\sigma_T,
  \label{eq:tau_baryon}
 \end{equation}
  where $r_i$ is the comoving radius of the jet, $n_e$ represents the comoving electron number density, which can be estimated by the comoving proton number density $n_p$ ($n_e$ = $n_p$, due to the electric neutrality for the jet) as:
 \begin{equation}
  n_p \approx \frac{L_{\rm iso}}{4\pi r_i^2c\Gamma^2m_pc^2}, \label{eq:n_p}
 \end{equation}
 where $L_{\text{iso}}$ is the isotropic luminosity and $m_p$ is the proton mass.

The saturated acceleration (i.e., when the photosphere radius exceeds the coasting radius) is considered to constrain the minimum baryon loading. In this case, the dimensionless entropy $\eta$, which indicates the average energy per baryon ($\eta$= $L/Mc^2$), should be equal to the bulk Lorentz factor ($\eta \simeq \Gamma$). According to equations \ref{eq:tau_baryon} and \ref{eq:n_p}, the photosphere radius ($\tau_b = 1$) can be estimated as
\begin{align}
 r_{\rm ph} = \frac{L_{\rm iso}\sigma_T}{4\pi c^3m_p\eta^3} = 1.9\times10^{15}\text{ cm}\bigg(\frac{L_{\rm iso}}{1.68\times10^{48} \text{ erg\ s}^{-1}}\bigg)\eta^{-3}\label{eq:rph},
\end{align}
where $L_{\text{iso}} \sim 1.68\times10^{48} \text{ erg\ s}^{-1}$ is taken from \citet{YJ2020}. To reconcile with our model, our estimated photosphere radius ($l_x \sim10^{6}$ cm) by equation (\ref{eq:E_mode}), which depends on the magnetic field strength and the pair density $n_{\pm}$, should be larger (thus dominated) than the above baryon-related $r_{\rm ph}$. This leads to a constraint on the lower limit of the $\eta$:
\begin{align}
 \eta \gtrsim 1248.4 \bigg(\frac{L_{\rm iso}}{ 1.68\times10^{48} \text{ erg\ s}^{-1}}\bigg)^{1/3}\bigg(\frac{r_{\rm ph}}{10^6 \text{ cm}}\bigg)^{-1/3}\label{eq:eta}.
\end{align}

Such a high $\eta$ therefore implies a very low level of baryon contamination. From equation (\ref{eq:n_p}), the proton number density can be constrained as:
\begin{align}
 n_p \lesssim 1.83\times10^{21}\text{ cm}^{-3}&\bigg(\frac{L_{\rm iso}}{1.68\times10^{48} \text{ erg\ s}^{-1}}\bigg)\bigg(\frac{r_{\rm ph}}{10^6 \text{ cm}}\bigg)^{-2}\nonumber \\
 &\times\bigg(\frac{\Gamma}{1248.4}\bigg)^{-2}. 
\end{align}

Furthermore, according to the formula of optical depth [equation (\ref{eq:E_mode})] in our model, the photosphere radius is also expected to increase due to the new electrons provided by the baryon contamination. These electrons ($n_e \sim n_p \lesssim 5.04\times10^{21}\text{ cm}^{-3}$), however, are insignificant in comparison to the $e^{\pm}$ number ($n_{\pm}\sim 4\times10^{23}\text{ cm}^{-3}$) provided by the magnetar wind. Thus, our model is not sensitive to the baryon loading.

\subsection{Magnetic Dissipation and Radiative Efficiency}
The extremely strong magnetic field of the magnetar invokes several effects during the MGF event. On one hand, near the magnetar surface, magnetic reconnection happens due to some magnetic instabilities. Then, electrons are accelerated to relativistic speed and a large number of seed photons are produced, dissipating the magnetic energy. On the other hand, above the magnetar surface, the strong magnetic field contributes to high scattering and absorption coefficients for the photons, causing a high opacity and thermalizing the photons (making a fireball). The trapped thermal emission can finally escape when the fireball expands to the photosphere radius ($\tau = 1$).

The energy from magnetic dissipation during the earlier period is all transferred to the thermal energy. Since extremely low baryon loading is considered, the jet mass (mainly from the $e^{\pm}$ pairs) is low. Then, the saturated-acceleration radius (where $\Gamma$=$L/Mc^2$) is quite large, and is well above the photosphere radius. Thus, the adiabatic cooling is insignificant. The observed efficiency of the thermal emission is almost 100$\%$.

According to our constrained value of the magnetic field ($B_{*}\sim 10^{16}$ G, see Table \ref{tab:MB_Wien}), the total magnetic energy of the magnetar can be estimated as
\begin{align}
E_{B} \simeq \frac{1}{6}B_*^2R_s^3 \simeq 1.67 \times10^{49}~{\rm erg}\ \bigg(\frac{B_{*}}{10^{16}~\rm G}\bigg)^{2}\bigg(\frac{R_{s}}{10^{6}~\rm cm}\bigg)^{3}. \label{eq:magE} 
\end{align}
The magnetization parameter $\sigma_0$ is defined as
\begin{align}
\sigma_0 = \frac{E_B}{E_{\rm iso}} = 1427.3\bigg(\frac{E_B}{1.67 \times10^{49}\text{ erg}}\bigg)\bigg(\frac{E_{\rm iso}}{1.17 \times10^{46}\text{ erg}}\bigg)^{-1}, \label{eq:sigma_0}
\end{align}
suggesting only a small fraction ($<1/1000$) of the total magnetic energy of the magnetar is converted to radiation.

Furthermore, we can estimate the local magnetic field energy $E_{B, \rm loc}$ using the initial size of the fireball bubble:
\begin{align}
E_{B, \rm loc} &\simeq \frac{1}{6}B_*^2l_0^3 \nonumber \\
&\simeq 1.67 \times10^{46}~{\rm erg}\ \bigg(\frac{B_{*}}{10^{16}~\rm G}\bigg)^{2}\bigg(\frac{l_0}{10^{5}~\rm cm}\bigg)^{3} \label{eq:LocB}. 
\end{align}
Noteworthily, the local energy of magnetic field $E_{B, \rm loc}$ is comparable to the observed $E_{\rm iso}$. This consistency supports that the radiative energy of MGF event is from the energy dissipation of the locally twisted magnetic field. A high dissipation efficiency of the local magnetic field can be calculated as 
\begin{equation}
    \varepsilon\sim E_{\rm iso}/E_{B, \rm loc} \sim 70 \%.
\end{equation}

\section{Summary}\label{sec:summary}
	
In this paper, we introduced a physical model which involves a fireball bubble emerging from the magnetar surface due to violent neutron star activities such as crust cracking or magnetic reconnection. We studied in detail the evolution of such a fireball, including its expansion and interaction with the magnetar $e^{\pm}$ wind. We found that high-energy photons at the photosphere radius are Comptonized by the high-number-density $e^{\pm}$ relativistic wind, which results in a MB spectrum with a Wien tail. Through a direct Monte-Carlo fit, our physical model is found well consistent with the observed spectra of GRB 200415A. Our results, for the first time, confirm the physical origin of the magnetar giant flares through first-principle theoretical calculations as well as the evidence of data consistency.

\section*{acknowledgments}
	
We thank the anonymous referee for constructive suggestions. We acknowledge the support by the National Key Research and Development Programs of China (2018YFA0404204, 2022YFF0711404, 2022SKA0130102), the National Natural Science Foundation of China (Grant Nos. 11833003, U2038105, 12121003), the science research grants from the China Manned Space Project with NO.CMS-CSST-2021-B11, and the Program for Innovative Talents. Y.Z.M. is supported by the National Postdoctoral Program for Innovative Talents (grant No. BX20200164). We acknowledge the use of public data from the Fermi Science Support Center (FSSC). We thank Di Xiao, Ken Chen, Qiao-Chu Li, Ze-Nan Liu, Jun Yang for helpful discussions and constructive comments. Z.J.Z. appreciates the help from Tianyue Li (School of Physics, NJU) and Shan Xiang (GXU) for their illustration support. 

\section*{Data Availability}
Upon reasonable requests, the code used to produce the figures and results in this paper is available from the corresponding authors. The {\it Fermi}/GBM data are publicly available at \url{https://heasarc.gsfc.nasa.gov/FTP/fermi/data}.

\clearpage	
	
\clearpage
\bibliographystyle{mnras}

\newpage
\appendix
\onecolumn
\section{SCATTERING OPACITY OF E-MODE AND O-MODE PHOTONS}\label{seca}
In this section, we list the analytical expressions of scattering opacity (see \citealt{HL2003MNRAS.338..233H} for detailed derivation) for E-mode ($\kappa^{\rm es}_1$) and O-mode ($\kappa^{\rm es}_2$). The expression of $\kappa^{\rm es}_1$ is as folows:
\begin{small}
\begin{align*}
 \text{\large{$\kappa^{\rm es}_1$}} = 730.769\left(\left(2.01 \times 10^{30} B^{2} n_{\pm}^{2} \cos \left[\theta_{B}\right]^{2}\left(1-\frac{8.06 \times 10^{7} n_{\pm}}{\nu^{2}}-\frac{1}{274 \pi}\times\right.\right.\right.
\end{align*}
\begin{align*}
 \left(1.272-\frac{1.36 \times 10^{27}}{B^{2}}-1.51 \times 10^{-14} B-\frac{4.4 \times 10^{13}\left(0.307+\ln \left[2.27 \times 10^{-14} B\right]\right)}{B}\right)+\frac{1}{274 \pi} 
\end{align*}
\begin{align*}
 \left.\left.\left(1.195-\frac{9.68 \times 10^{26}}{B^{2}}-\frac{2}{3} \ln \left[2.27 \times 10^{-14} B\right]-\frac{4.4 \times 10^{13}\left(0.855+\ln \left[2.27 \times 10^{-14} B\right]\right)}{B}\right)\right)^{2} \operatorname{sin}\left[\theta_{B}\right]^{2}\right) \Bigg/ 
\end{align*}
\begin{align*}
 \left(\left(1-\frac{7.83 \times 10^{12} B^{2}}{\nu^{2}}\right)^{2} \nu^{4}\left(9.34 \times 10^{-19} B^{4}+(2 \pi \nu)^{2}\right)\left(\left(\left(\frac{2}{3}-\frac{1.936 \times 10^{27}}{B^{2}}+\frac{4.4 \times 10^{13}\left(0.1447-\ln \left[2.27 \times 10^{-14} B\right]\right)}{B}\right)\right)\right.\right. 
\end{align*}
\begin{align*}
 \left(1-\frac{8.06 \times 10^{7} n_{\pm}}{\left(1-\frac{7.83 \times 10^{12} B^{2}}{\nu^{2}}\right) \nu^{2}}+\frac{1}{274 \pi}\left(1.195-\frac{9.68 \times 10^{26}}{B^{2}}-\frac{2}{3} \ln \left[2.27 \times 10^{-14} B\right]-\frac{4.4 \times 10^{13}\left(0.855+\ln \left[2.27 \times 10^{-14} B\right]\right)}{B}\right)\right) 
\end{align*}
\begin{align*}
 \left(1-\frac{8.06 \times 10^{7} n_{\pm}}{\nu^{2}}-\frac{1}{274 \pi}\left(1.272-\frac{1.36 \times 10^{27}}{B^{2}}-1.51 \times 10^{-14} B-\frac{4.4 \times 10^{13}\left(0.307+\ln \left[2.27 \times 10^{-14} B\right]\right)}{B}\right)+\right. 
\end{align*}
\begin{align*}
 \left.\left.\frac{1}{274 \pi}\left(1.195-\frac{9.68 \times 10^{26}}{B^{2}}-\frac{2}{3} \ln \left[2.27 \times 10^{-14} B\right]-\frac{4.4 \times 10^{13}\left(0.855+\ln \left[2.27 \times 10^{-14} B\right]\right)}{B}\right)\right) \operatorname{sin}\left[\theta_{B}\right]^{2}\right) \Bigg/ 
\end{align*}
\begin{align*}
 \left(274 \pi\left(1+\frac{1}{274 \pi}\left(1.195-\frac{9.68 \times 10^{26}}{B^{2}}-\frac{2}{3} \ln \left[2.27 \times 10^{-14} B\right]-\frac{4.4 \times 10^{13}\left(0.855+\ln \left[2.27 \times 10^{-14} B\right]\right)}{B}\right)\right)\right)+ 
\end{align*}
\begin{align*}
 \left(-1+\frac{5.08604 \times 10^{28} B^{2}n_{\pm}^{2}}{\left(1-\frac{7.83 \times 10^{12} B^{2}}{\nu^{2}}\right)^{2} \nu^{6}}+\frac{8.06 \times 10^{7}n_{\pm} }{\left(1-\frac{7.83 \times 10^{12} B^{2}}{\nu^{2}}\right) \nu^{2}}-\frac{1}{274 \pi}\left(1.195-\frac{9.68 \times 10^{26}}{B^{2}}-\frac{2}{3} \ln \left[2.27 \times 10^{-14} B\right]-\right.\right. 
\end{align*}
\begin{align*}
 \left.\frac{4.4 \times 10^{13}\left(0.855+\ln \left[2.27 \times 10^{-14} B\right]\right)}{B}\right)+\left(1-\frac{8.06 \times 10^{7} n_{\pm}}{\left(1-\frac{7.83 \times 10^{12} B^{2}}{\nu^{2}}\right) \nu^{2}}+\frac{1}{274 \pi}\right. 
\end{align*}
\begin{align*}
 \left.\left(1.195-\frac{9.68 \times 10^{26}}{B^{2}}-\frac{2}{3} \ln \left[2.27 \times 10^{-14} B\right]-\frac{4.4 \times 10^{13}\left(0.855+\ln \left[2.27 \times 10^{-14} B\right]\right)}{B}\right)\right) 
\end{align*}
\begin{align*}
 \left(1-\frac{8.06 \times 10^{7} n_{\pm}}{\nu^{2}}-\frac{1}{274 \pi}\left(1.272-\frac{1.36 \times 10^{27}}{B^{2}}-1.51 \times 10^{-14} B-\frac{4.4 \times 10^{13}\left(0.307+\ln \left[2.27 \times 10^{-14} B\right]\right)}{B}\right)+\right. 
\end{align*}
\begin{align*}
 \left.\left.\left.\frac{1}{274 \pi}\left(1.195-\frac{9.68 \times 10^{26}}{B^{2}}-\frac{2}{3} \ln \left[2.27 \times 10^{-14} B\right]-\frac{4.4 \times 10^{13}\left(0.855+\ln \left[2.27 \times 10^{-14} B\right]\right)}{B}\right)\right)\right) \operatorname{sin}\left[\theta_{B}\right]^{2}\right)^{2} 
\end{align*}
\begin{align*}
 \left(1+\left(5.08604 \times 10^{28} B^{2} n_{\pm}^{2} \operatorname{cos}\left[\theta_{B}\right]^{2}\left(1-\frac{8.06 \times 10^{7} n_{\pm}}{\nu^{2}}-\frac{1}{274 \pi}\left(1.272-\frac{1.36 \times 10^{27}}{B^{2}}-1.51 \times 10^{-14} B-\right.\right.\right. \right.
\end{align*}
\begin{align*}
 \left.\frac{4.4 \times 10^{13}\left(0.307+\ln \left[2.27 \times 10^{-14} B\right]\right)}{B}\right)+\frac{1}{274 \pi}\left(1.195-\frac{9.68 \times 10^{26}}{B^{2}}-\frac{2}{3} \ln \left[2.27 \times 10^{-14} B\right]-\right. 
\end{align*}
\begin{align*}
 \left.\left.\left.\frac{4.4 \times 10^{13}\left(0.855+\ln \left[2.27 \times 10^{-14} B\right]\right)}{B}\right)\right)^{2}\right) \Bigg/\left(\left(1-\frac{7.83 \times 10^{12} B^{2}}{\nu^{2}}\right)^{2} \nu^{6} \right.
\end{align*}
\begin{align*}
 \left(\left(\left(\frac{2}{3}-\frac{1.936 \times 10^{27}}{B^{2}}+\frac{4.4 \times 10^{13}\left(0.1447-\ln \left[2.27 \times 10^{-14} B\right]\right)}{B}\right)\left(1-\frac{8.06 \times 10^{7} n_{\pm}}{\left(1-\frac{7.83 \times 10^{12} B^{2}}{\nu^{2}}\right) \nu^{2}}+\frac{1}{274 \pi}\bigg(1.195-\right.\right.\right. 
\end{align*}
\begin{align*}
 \left.\left.\frac{9.68 \times 10^{26}}{B^{2}}-\frac{2}{3} \ln \left[2.27 \times 10^{-14} B\right]-\frac{4.4 \times 10^{13}\left(0.855+\ln \left[2.27 \times 10^{-14} B\right]\right)}{B}\right)\right)\left(1-\frac{8.06 \times 10^{7} n_{\pm}}{\nu^{2}}-\right. 
\end{align*} 
\begin{align*}
 \frac{1}{274 \pi}\left(1.272-\frac{1.36 \times 10^{27}}{B^{2}}-1.51 \times 10^{-14} B-\frac{4.4 \times 10^{13}\left(0.307+\ln \left[2.27 \times 10^{-14} B\right]\right)}{B}\right)+\frac{1}{274 \pi} 
\end{align*}
\begin{align*}
 \left.\left.\left(1.195-\frac{9.68 \times 10^{26}}{B^{2}}-\frac{2}{3} \ln \left[2.27 \times 10^{-14} B\right]-\frac{4.4 \times 10^{13}\left(0.855+\ln \left[2.27 \times 10^{-14} B\right]\right)}{B}\right)\right) \mathrm{sin}\left[\theta_{B}\right]^{2}\right) \Bigg/ 
\end{align*}
\begin{align*}
 \left(274 \pi\left(1+\frac{1}{274 \pi}\left(1.195-\frac{9.68 \times 10^{26}}{B^{2}}-\frac{2}{3} \ln \left[2.27 \times 10^{-14} B\right]-\frac{4.4 \times 10^{13}\left(0.855+\ln \left[2.27 \times 10^{-14} B\right]\right)}{B}\right)\right)+\right. 
\end{align*}
\begin{align*}
 \left(-1+\frac{5.08604 \times 10^{28} B^{2}n_{\pm}^2}{\left(1-\frac{7.83 \times 10^{12} B^{2}}{\nu^{2}}\right)^{2} \nu^{6}}+\frac{8.06 \times 10^{7} n_{\pm}}{\left(1-\frac{7.83 \times 10^{12} B^{2}}{\nu^{2}}\right) \nu^{2}}-\frac{1}{274 \pi}\left(1.195-\frac{9.68 \times 10^{26}}{B^{2}}-\frac{2}{3} \ln \left[2.27 \times 10^{-14} B\right]-\right.\right. 
\end{align*}
\begin{align*}
 \left.\frac{4.4 \times 10^{13}\left(0.855+\ln \left[2.27 \times 10^{-14} B\right]\right)}{B}\right)+\left(1-\frac{8.06 \times 10^{7} n_{\pm}}{\left(1-\frac{7.83 \times 10^{12} B^{2}}{\nu^{2}}\right) \nu^{2}}+\frac{1}{274 \pi}\left(1.195-\frac{9.68 \times 10^{26}}{B^{2}}-\right.\right. 
\end{align*}
\begin{align*}
 \left.\left.\frac{2}{3} \ln \left[2.27 \times 10^{-14} B\right]-\frac{4.4 \times 10^{13}\left(0.855+\ln \left[2.27 \times 10^{-14} B\right]\right)}{B}\right)\right)\left(1-\frac{8.06 \times 10^{7} n_{\pm}}{\nu^{2}}-\frac{1}{274 \pi}\right. 
\end{align*}
\begin{align*}
 \left(1.272-\frac{1.36 \times 10^{27}}{B^{2}}-1.51 \times 10^{-14} B-\frac{4.4 \times 10^{13}\left(0.307+\ln \left[2.27 \times 10^{-14} B\right]\right)}{B}\right)+\frac{1}{274 \pi} 
\end{align*}
\begin{align*}
 \left.\left.\left.\left.\left.\left(1.195-\frac{9.68 \times 10^{26}}{B^{2}}-\frac{2}{3} \ln \left[2.27 \times 10^{-14} B\right]-\frac{4.4 \times 10^{13}\left(0.855+\ln \left[2.27 \times 10^{-14} B\right]\right)}{B}\right)\right)\right)\left(\operatorname{sin}\left[\theta_{B}\right]^{2}\right)^{2}\right)\right)\right)+ 
\end{align*}
\begin{align*}
 \left(2 \pi^{2} \nu^{2}\left(1-\left(2.26 \times 10^{14} n_{\pm} \sqrt{\frac{B^{2}}{\nu^{2}}} \operatorname{cos}\left[\theta_{B}\right]^{2}\left(1-\frac{8.06 \times 10^{7} n_{\pm}}{\nu^{2}}-\frac{1}{274 \pi}\left(1.272-\frac{1.36 \times 10^{27}}{B^{2}}-1.51 \times 10^{-14} B-\right.\right.\right.\right.\right. 
\end{align*}
\begin{align*}
 \left.\frac{4.4 \times 10^{13}\left(0.307+\ln \left[2.27 \times 10^{-14} B\right]\right)}{B}\right)+\frac{1}{274 \pi}\left(1.195-\frac{9.68 \times 10^{26}}{B^{2}}-\frac{2}{3} \ln \left[2.27 \times 10^{-14} B\right]-\right. 
\end{align*}
\begin{align*}
 \left.\left.\left.\frac{4.4 \times 10^{13}\left(0.855+\ln \left[2.27 \times 10^{-14} B\right]\right)}{B}\right)\right)\right) \Bigg/\left(1-\frac{7.83 \times 10^{12} B^{2}}{\nu^{2}}\right) \nu^{2} 
\end{align*}
\begin{align*}
 \left(\left(\left(\frac{2}{3}-\frac{1.936 \times 10^{27}}{B^{2}}+\frac{4.4 \times 10^{13}\left(0.1447-\ln \left[2.27 \times 10^{-14} B\right]\right)}{B}\right)\left(1-\frac{8.06 \times 10^{7} n_{\pm}}{\left(1-\frac{7.83 \times 10^{12} B^{2}}{\nu^{2}}\right) \nu^{2}}+\frac{1}{274 \pi}\bigg(1.195-\right.\right.\right. 
\end{align*}
\begin{align*}
 \left.\left.\frac{9.68 \times 10^{26}}{B^{2}}-\frac{2}{3} \ln \left[2.27 \times 10^{-14} B\right]-\frac{4.4 \times 10^{13}\left(0.855+\ln \left[2.27 \times 10^{-14} B\right]\right)}{B}\right)\right)\left(1-\frac{8.06 \times 10^{7} n_{\pm}}{\nu^{2}}-\right. 
\end{align*}
\begin{align*}
 \frac{1}{274 \pi}\left(1.272-\frac{1.36 \times 10^{27}}{B^{2}}-1.51 \times 10^{-14} B-\frac{4.4 \times 10^{13}\left(0.307+\ln \left[2.27 \times 10^{-14} B\right]\right)}{B}\right)+\frac{1}{274 \pi} 
\end{align*}
\begin{align*}
 \left.\left.\left(1.195-\frac{9.68 \times 10^{26}}{B^{2}}-\frac{2}{3} \ln \left[2.27 \times 10^{-14} B\right]-\frac{4.4 \times 10^{13}\left(0.855+\ln \left[2.27 \times 10^{-14} B\right]\right)}{B}\right)\right) \operatorname{sin}\left[\theta_{B}\right]^{2}\right) \Bigg/ 
\end{align*}
\begin{align*}
 \left(274 \pi\left(1+\frac{1}{274 \pi}\left(1.195-\frac{9.68 \times 10^{26}}{B^{2}}-\frac{2}{3} \ln \left[2.27 \times 10^{-14} B\right]-\frac{4.4 \times 10^{13}\left(0.855+\ln \left[2.27 \times 10^{-14} B\right]\right)}{B}\right)\right)\right)+ 
\end{align*}
\begin{align*}
 \left(-1+\frac{5.08604 \times 10^{28} B^{2} n_{\pm}^{2}}{\left(1-\frac{7.83 \times 10^{12} B^{2}}{\nu^{2}}\right)^{2} \nu^{6}}+\frac{8.06 \times 10^{7} n_{\pm}}{\left(1-\frac{7.83 \times 10^{12} B^{2}}{\nu^{2}}\right) \nu^{2}}-\frac{1}{274 \pi}\left(1.195-\frac{9.68 \times 10^{26}}{B^{2}}-\frac{2}{3} \ln \left[2.27 \times 10^{-14} B\right]-\right.\right. 
\end{align*}
\begin{align*}
 \left.\frac{4.4 \times 10^{13}\left(0.855+\ln \left[2.27 \times 10^{-14} B\right]\right)}{B}\right)+\left(1-\frac{8.06 \times 10^{7} n_{\pm}}{\left(1-\frac{7.83 \times 10^{12} B^{2}}{\nu^{2}}\right) \nu^{2}}+\frac{1}{274 \pi}\left(1.195-\frac{9.68 \times 10^{26}}{B^{2}}-\right.\right. 
\end{align*}
\begin{align*}
 \left.\left.\frac{2}{3} \ln \left[2.27 \times 10^{-14} B\right]-\frac{4.4 \times 10^{13}\left(0.855+\ln \left[2.27 \times 10^{-14} B\right]\right)}{B}\right)\right)\left(1-\frac{8.06 \times 10^{7} n_{\pm}}{\nu^{2}}-\frac{1}{274 \pi}\right. 
\end{align*}
\begin{align*}
 \left(1.272-\frac{1.36 \times 10^{27}}{B^{2}}-1.51 \times 10^{-14} B-\frac{4.4 \times 10^{13}\left(0.307+\ln \left[2.27 \times 10^{-14} B\right]\right)}{B}\right)+\frac{1}{274 \pi}\Bigg(1.195- 
\end{align*}
\begin{align*}
\left.\left.\left.\left.\left.\left.\frac{9.68 \times 10^{26}}{B^{2}}-\frac{2}{3} \ln \left[2.27 \times 10^{-14} B\right]-\frac{4.4 \times 10^{13}\left(0.855+\ln \left[2.27 \times 10^{-14} B\right]\right)}{B}\right)\right)\operatorname{sin}\left[\theta_{B}\right]^{2}\right)\right)\right)^{2} \right) \Bigg/ 
\end{align*}
\begin{align*}
 \left(\left(9.34 \times 10^{-19} B^{4}+\left(1.75824 \times 10^{7} B+2 \pi \nu\right)^{2}\right)\left(1+\left(5.08604 \times 10^{28} B^{2} n_{\pm}^{2} \operatorname{cos}\left[\theta_{B}\right]^{2}\left(1-\frac{8.06 \times 10^{7} n_{\pm}}{\nu^{2}}-\frac{1}{274 \pi}\right.\right.\right.\right. 
\end{align*}
\begin{align*}
 \left(1.272-\frac{1.36 \times 10^{27}}{B^{2}}-1.51 \times 10^{-14} B-\frac{4.4 \times 10^{13}\left(0.307+\ln \left[2.27 \times 10^{-14} B\right]\right)}{B}\right)+\frac{1}{274 \pi}\left(1.195-\frac{9.68 \times 10^{26}}{B^{2}}-\right. 
\end{align*}
\begin{align*}
 \left.\left.\left.\frac{2}{3} \ln \left[2.27 \times 10^{-14} B\right]-\frac{4.4 \times 10^{13}\left(0.855+\ln \left[2.27 \times 10^{-14} B\right]\right)}{B}\right)\right)^{2}\right) \Bigg/\left(1-\frac{7.83 \times 10^{12} B^{2}}{\nu^{2}}\right)^{2} \nu^{6} 
\end{align*}
\begin{align*}
 \left(1.272-\frac{1.36 \times 10^{27}}{B^{2}}-1.51 \times 10^{-14} B-\frac{4.4 \times 10^{13}\left(0.307+\ln \left[2.27 \times 10^{-14} B\right]\right)}{B}\right)+\frac{1}{274 \pi}\Bigg(1.195- 
\end{align*}
\begin{align*}
\left.\left.\left.\left.\left.\left.\left.\frac{9.68 \times 10^{26}}{B^{2}}-\frac{2}{3} \ln \left[2.27 \times 10^{-14} B\right]-\frac{4.4 \times 10^{13}\left(0.855+\ln \left[2.27 \times 10^{-14} B\right]\right)}{B}\right) \right)\right) \operatorname{sin}^{2}\left[\theta_{B}\right]^{2}\right)\right) \right)^{2}\right) \Bigg/ 
\end{align*}
\begin{align*}
 \left(\left(9.34 \times 10^{-19} B^{4}+\left(1.75824 \times 10^{7} B+2 \pi \nu\right)^{2}\right)\left(1+\left(5.08604 \times 10^{28} B^{2} n_{\pm}^{2} \cos \left[\theta_{B}\right]^{2}\left(1-\frac{8.06 \times 10^{7} n_{\pm}}{\nu^{2}}-\frac{1}{274 \pi}\right.\right.\right.\right. 
\end{align*}
\begin{align*}
 \left(1.272-\frac{1.36 \times 10^{27}}{B^{2}}-1.51 \times 10^{-14} B-\frac{4.4 \times 10^{13}\left(0.307+\ln \left[2.27 \times 10^{-14} B\right]\right)}{B}\right)+\frac{1}{274 \pi}\left(1.195-\frac{9.68 \times 10^{26}}{B^{2}}-\right. 
\end{align*}
\begin{align*}
 \left.\left.\left.\frac{2}{3} \ln \left[2.27 \times 10^{-14} B\right]-\frac{4.4 \times 10^{13}\left(0.855+\ln \left[2.27 \times 10^{-14} B\right]\right)}{B}\right)\right)^{2}\right) \Bigg/\left(1-\frac{7.83 \times 10^{12} B^{2}}{\nu^{2}}\right)^{2} \nu^{6} 
\end{align*}
\begin{align*}
 \left(\left(\left(\frac{2}{3}-\frac{1.936 \times 10^{27}}{B^{2}}+\frac{4.4 \times 10^{13}\left(0.1447-\ln \left[2.27 \times 10^{-14} B\right]\right)}{B}\right)\left(1-\frac{8.06 \times 10^{7} n_{\pm}}{\left(1-\frac{7.83 \times 10^{12} B^{2}}{\nu^{2}}\right) \nu^{2}}+\frac{1}{274 \pi}\Bigg(1.195-\right.\right.\right. 
\end{align*}
\begin{align*}
 \left.\left.\frac{9.68 \times 10^{26}}{B^{2}}-\frac{2}{3} \ln \left[2.27 \times 10^{-14} B\right]-\frac{4.4 \times 10^{13}\left(0.855+\ln \left[2.27 \times 10^{-14} B\right]\right)}{B}\right)\right)\left(1-\frac{8.06 \times 10^{7} n_{\pm}}{\nu^{2}}-\right. 
\end{align*}
\begin{align*}
 \frac{1}{274 \pi}\left(1.272-\frac{1.36 \times 10^{27}}{B^{2}}-1.51 \times 10^{-14} B-\frac{4.4 \times 10^{13}\left(0.307+\ln \left[2.27 \times 10^{-14} B\right]\right)}{B}\right)+\frac{1}{274 \pi} 
\end{align*}
\begin{align*}
 \left.\left.\left(1.195-\frac{9.68 \times 10^{26}}{B^{2}}-\frac{2}{3} \ln \left[2.27 \times 10^{-14} B\right]-\frac{4.4 \times 10^{13}\left(0.855+\ln \left[2.27 \times 10^{-14} B\right]\right)}{B}\right)\right) \operatorname{sin}\left[\theta_{B}\right]^{2}\right) \Bigg/ 
\end{align*}
\begin{align*}
 \left.\left(274 \pi\left(1+\frac{1}{274 \pi}\left(1.195-\frac{9.68 \times 10^{26}}{B^{2}}-\frac{2}{3} \ln \left[2.27 \times 10^{-14} B\right]-\frac{4.4 \times 10^{13}\left(0.855+\ln \left[2.27 \times 10^{-14} B\right]\right)}{B}\right)\right)\right)+\right. 
\end{align*}
\begin{align*}
 \left(-1+\frac{5.08604 \times 10^{28} B^{2} n_{\pm}^{2}}{\left(1-\frac{7.83 \times 10^{12} B^{2}}{\nu^{2}}\right)^{2} \nu^{6}}+\frac{8.06 \times 10^{7} n_{\pm}}{\left(1-\frac{7.83 \times 10^{12} B^{2}}{\nu^{2}}\right) \nu^{2}}-\frac{1}{274 \pi}\left(1.195-\frac{9.68 \times 10^{26}}{B^{2}}-\frac{2}{3} \mathrm{Log}\left[2.27 \times 10^{-14} B\right]-\right.\right. 
\end{align*}
\begin{align*}
 \left.\frac{4.4 \times 10^{13}\left(0.855+\ln \left[2.27 \times 10^{-14} B\right]\right)}{B}\right)+\left(1-\frac{8.06 \times 10^{7} n_{\pm}}{\left(1-\frac{7.83 \times 10^{12} B^{2}}{\nu^{2}}\right) \nu^{2}}+\frac{1}{274 \pi}\left(1.195-\frac{9.68 \times 10^{26}}{B^{2}}-\right.\right. 
\end{align*}
\begin{align*}
 \left.\left.\frac{2}{3} \ln \left[2.27 \times 10^{-14} B\right]-\frac{4.4 \times 10^{13}\left(0.855+\ln \left[2.27 \times 10^{-14} B\right]\right)}{B}\right)\right)\left(1-\frac{8.06 \times 10^{7} n_{\pm}}{\nu^{2}}-\frac{1}{274 \pi}\right. 
\end{align*}
\begin{align*}
 \left(1.272-\frac{1.36 \times 10^{27}}{B^{2}}-1.51 \times 10^{-14} B-\frac{4.4 \times 10^{13}\left(0.307+\ln \left[2.27 \times 10^{-14} B\right]\right)}{B}\right)+\frac{1}{274 \pi} 
\end{align*}
\begin{align*}
 \left.\left.\left.\left.\left.\left.\left(1.195-\frac{9.68 \times 10^{26}}{B^{2}}-\frac{2}{3} \ln \left[2.27 \times 10^{-14} B\right]-\frac{4.4 \times 10^{13}\left(0.855+\ln \left[2.27 \times 10^{-14} B\right]\right)}{B}\right)\right)\right)\operatorname{sin}\left[\theta_{B}\right]^{2}\right)^{2}\right)\right)\right)+ 
\end{align*}
\begin{align*}
 \left(2 \pi^{2} \nu^{2}\left(1+\left(2.26 \times 10^{14}n_{\pm} \sqrt{\frac{B^{2}}{\nu^{2}}} \operatorname{cos}\left[\theta_{B}\right]^{2}\left(1-\frac{8.06 \times 10^{7}n_{\pm}}{\nu^{2}}-\frac{1}{274 \pi}\left(1.272-\frac{1.36 \times 10^{27}}{B^{2}}-1.51 \times 10^{-14} B-\right.\right.\right.\right.\right. 
\end{align*}
\begin{align*}
 \left.\frac{4.4 \times 10^{13}\left(0.307+\ln \left[2.27 \times 10^{-14} B\right]\right)}{B}\right)+\frac{1}{274 \pi}\left(1.195-\frac{9.68 \times 10^{26}}{B^{2}}-\frac{2}{3} \ln \left[2.27 \times 10^{-14} B\right]-\right. 
\end{align*}
\begin{align*}
 \left.\left.\frac{4.4 \times 10^{13}\left(0.855+\ln \left[2.27 \times 10^{-14} B\right]\right)}{B}\right)\right) \Bigg/\left(\left(1-\frac{7.83 \times 10^{12} B^{2}}{\nu^{2}}\right) \nu^{2}\right. 
\end{align*}
\begin{align*}
 \left(\left(\left(\frac{2}{3}-\frac{1.936 \times 10^{27}}{B^{2}}+\frac{4.4 \times 10^{13}\left(0.1447-\ln \left[2.27 \times 10^{-14} B\right]\right)}{B}\right)\left(1-\frac{8.06 \times 10^{7} n_{\pm}}{\left(1-\frac{7.83 \times 10^{12} B^{2}}{\nu^{2}}\right) \nu^{2}}+\frac{1}{274 \pi}\Bigg(1.195-\right.\right.\right. 
\end{align*}
\begin{align*}
 \left.\left.\frac{9.68 \times 10^{26}}{B^{2}}-\frac{2}{3} \ln \left[2.27 \times 10^{-14} B\right]-\frac{4.4 \times 10^{13}\left(0.855+\ln \left[2.27 \times 10^{-14} B\right]\right)}{B}\right)\right)\left(1-\frac{8.06 \times 10^{7} n_{\pm}}{\nu^{2}}-\right. 
\end{align*}
\begin{align*}
 \frac{1}{274 \pi}\left(1.272-\frac{1.36 \times 10^{27}}{B^{2}}-1.51 \times 10^{-14} B-\frac{4.4 \times 10^{13}\left(0.307+\ln \left[2.27 \times 10^{-14} B\right]\right)}{B}\right)+\frac{1}{274 \pi} 
\end{align*}
\begin{align*}
 \left.\left.\left(1.195-\frac{9.68 \times 10^{26}}{B^{2}}-\frac{2}{3} \ln \left[2.27 \times 10^{-14} B\right]-\frac{4.4 \times 10^{13}\left(0.855+\ln \left[2.27 \times 10^{-14} B\right]\right)}{B}\right)\right) \operatorname{sin}\left[\theta_{B}\right]^{2}\right) \Bigg/ 
\end{align*}
\begin{align*}
 \left.\left(274 \pi\left(1+\frac{1}{274 \pi}\left(1.195-\frac{9.68 \times 10^{26}}{B^{2}}-\frac{2}{3} \ln \left[2.27 \times 10^{-14} B\right]-\frac{4.4 \times 10^{13}\left(0.855+\ln \left[2.27 \times 10^{-14} B\right]\right)}{B}\right)\right)\right)+\right. 
\end{align*}
\begin{align*}
 \left(-1+\frac{5.08604 \times 10^{28} B^{2} n_{\pm}^{2}}{\left(1-\frac{7.83 \times 10^{12} B^{2}}{\nu^{2}}\right)^{2} \nu^{6}}+\frac{8.06 \times 10^{7} n_{\pm}}{\left(1-\frac{7.83 \times 10^{12} B^{2}}{\nu^{2}}\right) \nu^{2}}-\frac{1}{274 \pi}\left(1.195-\frac{9.68 \times 10^{26}}{B^{2}}-\frac{2}{3} \ln\left[2.27 \times 10^{-14} B\right]-\right.\right. 
\end{align*}
\begin{align*}
 \left.\frac{4.4 \times 10^{13}\left(0.855+\ln \left[2.27 \times 10^{-14} B\right]\right)}{B}\right)+\left(1-\frac{8.06 \times 10^{7} n_{\pm}}{\left(1-\frac{7.83 \times 10^{12} B^{2}}{\nu^{2}}\right) \nu^{2}}+\frac{1}{274 \pi}\left(1.195-\frac{9.68 \times 10^{26}}{B^{2}}-\right.\right. 
\end{align*}
\begin{align*}
 \left.\left.\frac{2}{3} \ln \left[2.27 \times 10^{-14} B\right]-\frac{4.4 \times 10^{13}\left(0.855+\ln \left[2.27 \times 10^{-14} B\right]\right)}{B}\right)\right)\left(1-\frac{8.06 \times 10^{7} n_{\pm}}{\nu^{2}}-\frac{1}{274 \pi}\right. 
\end{align*}
\begin{align*}
 \left(1.272-\frac{1.36 \times 10^{27}}{B^{2}}-1.51 \times 10^{-14} B-\frac{4.4 \times 10^{13}\left(0.307+\ln \left[2.27 \times 10^{-14} B\right]\right)}{B}\right)+\frac{1}{274 \pi}\Bigg(1.195- 
\end{align*}
\begin{align*}
\left.\left.\left. \left.\left.\left.\left.\frac{9.68 \times 10^{26}}{B^{2}}-\frac{2}{3} \ln \left[2.27 \times 10^{-14} B\right]-\frac{4.4 \times 10^{13}\left(0.855+\ln \left[2.27 \times 10^{-14} B\right]\right)}{B}\right)\right)\right)\operatorname{sin}\left[\theta_{B}\right]^{2}\right) \right)\right)^{2}\right) \Bigg/ 
\end{align*}

\begin{align*}
 \left(\left(9.34 \times 10^{-19} B^{4}+\left(-1.75824 \times 10^{7} B+2 \pi \nu\right)^{2}\right)\left(1+\left(5.08604 \times 10^{28} B^{2} n_{\pm}^{2} \cos \left[\theta_{B}\right]^{2}\left(1-\frac{8.06 \times 10^{7} n_{\pm}}{\nu^{2}}-\frac{1}{274 \pi}\right.\right.\right.\right. 
\end{align*}
\begin{align*}
 \left(1.272-\frac{1.36 \times 10^{27}}{B^{2}}-1.51 \times 10^{-14} B-\frac{4.4 \times 10^{13}\left(0.307+\ln \left[2.27 \times 10^{-14} B\right]\right)}{B}\right)+\frac{1}{274 \pi}\times
\end{align*}
\begin{align*}
 \left.\left.\left(1.195-\frac{9.68 \times 10^{26}}{B^{2}}-\frac{2}{3} \ln \left[2.27 \times 10^{-14} B\right]-\frac{4.4 \times 10^{13}\left(0.855+\ln \left[2.27 \times 10^{-14} B\right]\right)}{B}\right)\right)^{2}\right)\Bigg/ 
\end{align*}
\begin{align*}
 \left(\left(1-\frac{7.83 \times 10^{12} B^{2}}{\nu^{2}}\right)^{2} \nu^{6}\left(\left(\left(\frac{2}{3}-\frac{1.936 \times 10^{27}}{B^{2}}+\frac{4.4 \times 10^{13}\left(0.1447-\ln \left[2.27 \times 10^{-14} B\right]\right)}{B}\right)\right.\right.\right. 
\end{align*}
\begin{align*}
 \left(1-\frac{8.06 \times 10^{7} n_{\pm}}{\left(1-\frac{7.83 \times 10^{12} B^{2}}{\nu^{2}}\right) \nu^{2}}+\frac{1}{274 \pi}\left(1.195-\frac{9.68 \times 10^{26}}{B^{2}}-\frac{2}{3} \ln \left[2.27 \times 10^{-14} B\right]-\right.\right. 
\end{align*}
\begin{align*}
 \left.\left.\frac{4.4 \times 10^{13}\left(0.855+\ln \left[2.27 \times 10^{-14} B\right]\right)}{B}\right)\right)\left(1-\frac{8.06 \times 10^{7} n_{\pm}}{\nu^{2}}-\frac{1}{274 \pi}\right. 
\end{align*}
\begin{align*}
 \left(1.272-\frac{1.36 \times 10^{27}}{B^{2}}-1.51 \times 10^{-14} B-\frac{4.4 \times 10^{13}\left(0.307+\ln \left[2.27 \times 10^{-14} B\right]\right)}{B}\right)+\frac{1}{274 \pi} 
\end{align*}
\begin{align*}
 \left.\left.\left(1.195-\frac{9.68 \times 10^{26}}{B^{2}}-\frac{2}{3} \ln \left[2.27 \times 10^{-14} B\right]-\frac{4.4 \times 10^{13}\left(0.855+\ln \left[2.27 \times 10^{-14} B\right]\right)}{B}\right)\right) \operatorname{sin}\left[\theta_{B}\right]^{2}\right) \Bigg/ 
\end{align*}
\begin{align*}
 \left(274 \pi\left(1+\frac{1}{274 \pi}\left(1.195-\frac{9.68 \times 10^{26}}{B^{2}}-\frac{2}{3} \ln \left[2.27 \times 10^{-14} B\right]-\frac{4.4 \times 10^{13}\left(0.855+\ln \left[2.27 \times 10^{-14} B\right]\right)}{B}\right)\right)\right)+ 
\end{align*}
\begin{align*}
 \left(-1+\frac{5.08604 \times 10^{28} B^{2} n_{\pm}^{2}}{\left(1-\frac{7.83 \times 10^{12} B^{2}}{\nu^{2}}\right)^{2} \nu^{6}}+\frac{8.06 \times 10^{7} n_{\pm}}{\left(1-\frac{7.83 \times 10^{12} B^{2}}{\nu^{2}}\right) \nu^{2}}-\frac{1}{274 \pi}\left(1.195-\frac{9.68 \times 10^{26}}{B^{2}}-\frac{2}{3} \ln \left[2.27 \times 10^{-14} B\right]-\right.\right. 
\end{align*}
\begin{align*}
 \left.\frac{4.4 \times 10^{13}\left(0.855+\ln \left[2.27 \times 10^{-14} B\right]\right)}{B}\right)+\left(1-\frac{8.06 \times 10^{7} n_{\pm}}{\left(1-\frac{7.83 \times 10^{12} B^{2}}{\nu^{2}}\right) \nu^{2}}+\frac{1}{274 \pi}\left(1.195-\frac{9.68 \times 10^{26}}{B^{2}}-\right.\right. 
\end{align*}
\begin{align*}
 \left.\left.\frac{2}{3} \ln \left[2.27 \times 10^{-14} B\right]-\frac{4.4 \times 10^{13}\left(0.855+\ln \left[2.27 \times 10^{-14} B\right]\right)}{B}\right)\right)\left(1-\frac{8.06 \times 10^{7} n_{\pm}}{\nu^{2}}-\frac{1}{274 \pi}\right. 
\end{align*}
\begin{align*}
 \left(1.272-\frac{1.36 \times 10^{27}}{B^{2}}-1.51 \times 10^{-14} B-\frac{4.4 \times 10^{13}\left(0.307+\ln \left[2.27 \times 10^{-14} B\right]\right)}{B}\right)+\frac{1}{274 \pi} 
\end{align*}
\begin{align}
\left.\left. \left.\left.\left.\left.\left(1.195-\frac{9.68 \times 10^{26}}{B^{2}}-\frac{2}{3} \ln \left[2.27 \times 10^{-14} B\right]-\frac{4.4 \times 10^{13}\left(0.855+\ln \left[2.27 \times 10^{-14} B\right]\right)}{B}\right)\right)\operatorname{sin}\left[\theta_{B}\right]^{2}\right)^{2}\right)\right)\right)\right).
\label{eq:ap1}
\end{align}
\end{small}
\\
\\
\\

The expression of $\kappa^{\rm es}_2$ is as folows:

\begin{small}
\begin{align*}
 \text{\large{$\kappa^{\rm es}_2$}} = 730.769\left(\left(2 \pi \nu^{2}\left(1-\left(4.43415 \times 10^{-15}\left(1-\frac{7.83864 \times 16^{12} 8^{2}}{\nu^{2}}\right) \nu^{2}\right.\right.\right.\right. 
\end{align*}
\begin{align*}
 \left(\left(\left(\frac{2}{3}-\frac{1.936 \times 10^{27}}{B^{2}}+\frac{4.4 \times 10^{13}\left(0.1447-\ln \left[2.27 \times 10^{-14} B\right]\right)}{B}\right)\left(1-\frac{8.06 \times 10^{7} n_{\pm}}{\left(1-\frac{7.83 \times 10^{12} B^{2}}{\nu^{2}}\right) \nu^{2}}+\frac{1}{274 \pi}\Bigg(1.195-\right.\right.\right. 
\end{align*}
\begin{align*}
 \left.\left.\frac{9.68 \times 10^{26}}{B^{2}}-\frac{2}{3} \ln \left[2.27 \times 10^{-14} B\right]-\frac{4.4 \times 10^{13}\left(0.855+\ln \left[2.27 \times 10^{-14} B\right]\right)}{B}\right)\right)\left(1-\frac{8.06 \times 10^{7} n_{\pm}}{\nu^{2}}-\right. 
\end{align*}
\begin{align*}
 \frac{1}{274 \pi}\left(1.272-\frac{1.356 \times 10^{27}}{B^{2}}-1.51515 \times 10^{-14} B-\frac{4.4 \times 10^{13}\left(0.307+\ln \left[2.27 \times 10^{-14} B\right]\right)}{B}\right)+\frac{1}{274 \pi} 
\end{align*}
\begin{align*}
 \left.\left.\left(1.195-\frac{9.68 \times 10^{26}}{B^{2}}-\frac{2}{3} \ln \left[2.27 \times 10^{-14} B\right]-\frac{4.4 \times 10^{13}\left(0.855+\ln \left[2.27 \times 10^{-14} B\right]\right)}{B}\right)\right) \operatorname{sin}\left[\theta_{B}\right]^{2}\right) \Bigg/ 
\end{align*}
\begin{align*}
 \left(274 \pi\left(1+\frac{1}{274 \pi}\left(1.195-\frac{9.68 \times 10^{26}}{B^{2}}-\frac{2}{3} \ln \left[2.27 \times 10^{-14} B\right]-\frac{4.4 \times 10^{13}\left(0.855+\ln \left[2.27 \times 10^{-14} B\right]\right)}{B}\right)\right)\right)+ 
\end{align*}
\begin{align*}
 \left(-1+\frac{5.086 \times 10^{28} B^{2} n_{\pm}^{2}}{\left(1-\frac{7.83 \times 10^{12} B^{2}}{\nu^{2}}\right)^{2} \nu^{6}}+\frac{8.06 \times 10^{7} n_{\pm}}{\left(1-\frac{7.83 \times 10^{12} B^{2}}{\nu^{2}}\right) \nu^{2}}-\frac{1}{274 \pi}\left(1.195-\frac{9.68 \times 10^{26}}{B^{2}}-\frac{2}{3} \ln \left[2.27 \times 10^{-14} B\right]-\right.\right. 
\end{align*}
\begin{align*}
 \left.\frac{4.4 \times 10^{13}\left(0.855+\ln \left[2.27 \times 10^{-14} B\right]\right)}{B}\right)+\left(1-\frac{8.06 \times 10^{7} n_{\pm}}{\left(1-\frac{7.83 \times 10^{12} B^{2}}{\nu^{2}}\right) \nu^{2}}+\frac{1}{274 \pi}\left(1.195-\frac{9.68 \times 10^{26}}{B^{2}}-\right.\right. 
\end{align*}
\begin{align*}
\left.\left.\frac{2}{3} \ln \left[2.27 \times 10^{-14} B\right]-\frac{4.4 \times 10^{13}\left(0.855+\ln \left[2.27 \times 10^{-14} B\right]\right)}{B}\right)\right)\left(1-\frac{8.06 \times 10^{7} n_{\pm}}{\nu^{2}}-\frac{1}{274 \pi}\right. 
\end{align*}
\begin{align*}
 \left(1.272-\frac{1.356 \times 10^{27}}{B^{2}}-1.51515 \times 10^{-14} B-\frac{4.4 \times 10^{13}\left(0.307+\ln \left[2.27 \times 10^{-14} B\right]\right)}{B}\right)+\frac{1}{274 \pi} 
\end{align*}
\begin{align*}
 \left.\left.\left.\left.\left(1.195-\frac{9.68 \times 10^{26}}{B^{2}}-\frac{2}{3} \ln \left[2.27 \times 10^{-14} B\right]-\frac{4.4 \times 10^{13}\left(0.855+\ln \left[2.27 \times 10^{-14} B\right]\right)}{B}\right)\right)\right) \mid \operatorname{sin}\left[\theta_{B}\right]^{2}\right)\right) \Bigg/ 
\end{align*}
\begin{align*}
 \left(n_{\pm} \sqrt{\frac{B^{2}}{\nu^{2}}}\left(1-\frac{8.06 \times 10^{7} n_{\pm}}{\nu^{2}}-\frac{1}{274 \pi}\left(1.272-\frac{1.356 \times 10^{27}}{B^{2}}-1.51515 \times 10^{-14} B-\right.\right.\right. 
\end{align*}
\begin{align*}
 \left.\frac{4.4 \times 10^{13}\left(0.307+\ln \left[2.27 \times 10^{-14} B\right]\right)}{B}\right)+\frac{1}{274 \pi}\left(1.195-\frac{9.68 \times 10^{26}}{B^{2}}-\right. 
\end{align*}
\begin{align*}
\left.\left.\left.\left.\left.\frac{2}{3} \ln \left[2.27 \times 10^{-14} B\right]-\frac{4.4 \times 10^{13}\left(0.855+\ln \left[2.27 \times 10^{-14} B\right]\right)}{B}\right)\right) \right)\right)^{2}\right) \Bigg/ 
\end{align*}
\begin{align*}
 \left(\left(9.34 \times 10^{-19} B^{4}+\left(-1.76 \times 10^{7} B+2 \pi \nu\right)^{2}\right)\left(1+\left(1.96617 \times 10^{-29}\left(1-\frac{7.83 \times 10^{12} B^{2}}{\nu^{2}}\right)^{2} \nu^{6} \operatorname{sec}\left[\theta_{B}\right]^{2}\right.\right.\right. 
\end{align*}
\begin{align*}
 \left(\left(\left(\frac{2}{3}-\frac{1.936 \times 10^{27}}{B^{2}}+\frac{4.4 \times 10^{13}\left(0.1447-\ln \left[2.27 \times 10^{-14} B\right]\right)}{B}\right)\right.\right. 
\end{align*}
\begin{align*}
 \left(1-\frac{8.06 \times 10^{7} n_{\pm}}{\left(1-\frac{7.83 \times 10^{12} B^{2}}{\nu^{2}}\right) \nu^{2}}+\frac{1.195-\frac{9.68 \times 10^{26}}{B^{2}}-\frac{2}{3} \ln \left[2.27 \times 10^{-14} B\right]-\frac{4.4 \times 10^{13}\left(0.855+\ln \left[2.27 \times 10^{-14} B\right]\right)}{B}}{274 \pi}\right) 
\end{align*}
\begin{align*}
 \left(1-\frac{8.06 \times 10^{7} n_{\pm}}{\nu^{2}}-\frac{1.272-\frac{1.356 \times 10^{27}}{B^{2}}-1.51515 \times 10^{-14} B-\frac{4.4 \times 10^{13}\left(0.307+\ln \left[2.27 \times 10^{-14} B\right]\right)}{B}}{274 \pi}+\right. 
\end{align*}
\begin{align*}
 \left.\left.\frac{1.195-\frac{9.68 \times 10^{26}}{B^{2}}-\frac{2}{3} \ln \left[2.27 \times 10^{-14} B\right]-\frac{4.4 \times 10^{13}\left(0.855+\ln \left[2.27 \times 10^{-14} B\right]\right)}{B}}{274 \pi}\right) \operatorname{sin}\left[\theta_{B}\right]^{2}\right) \Bigg/\Bigg(274 \pi
\end{align*}
\begin{align*}
 \left.\left(1+\frac{1.195-\frac{9.68 \times 10^{26}}{B^{2}}-\frac{2}{3} \ln \left[2.27 \times 10^{-14} B\right]-\frac{4.4 \times 10^{13}\left(0.855+\ln \left[2.27 \times 10^{-14} B\right]\right)}{B}}{274 \pi}\right)\right)+\left(-1+\frac{5.086 \times 10^{28} B^{2} n_{\pm}^{2}}{\left(1-\frac{7.83 \times 10^{12} B^{2}}{\nu^{2}}\right)^{2} \nu^{6}}+\right. 
\end{align*}
\begin{align*}
 \frac{8.06 \times 10^{7} n_{\pm}}{\left(1-\frac{7.83 \times 10^{12} B^{2}}{\nu^{2}}\right) \nu^{2}}-\frac{1.195-\frac{9.68 \times 10^{26}}{B^{2}}-\frac{2}{3} \ln \left[2.27 \times 10^{-14} B\right]-\frac{4.4 \times 10^{13}\left(0.855+\ln \left[2.27 \times 10^{-14} B\right]\right)}{B}}{274 \pi}+ 
\end{align*}
\begin{align*}
 \left(1-\frac{8.06 \times 10^{7} n_{\pm}}{\left(1-\frac{7.83 \times 10^{12} B^{2}}{\nu^{2}}\right) \nu^{2}}+\frac{1.195-\frac{9.68 \times 10^{26}}{B^{2}}-\frac{2}{3} \ln \left[2.27 \times 10^{-14} B\right]-\frac{4.4 \times 10^{13}\left(0.855+\ln \left[2.27 \times 10^{-14} B\right]\right)}{B}}{274 \pi}\right) 
\end{align*}
\begin{align*}
 \left(1-\frac{8.06 \times 10^{7} n_{\pm}}{\nu^{2}}-\frac{1.272-\frac{1.356 \times 10^{27}}{B^{2}}-1.51515 \times 10^{-14} B-\frac{4.4 \times 10^{13}\left(0.307+\ln \left[2.27 \times 10^{-14} B\right]\right)}{B}}{274 \pi}+\right. 
\end{align*}
\begin{align*}
 \left.\left.\left.\frac{1.195-\frac{9.68 \times 10^{26}}{B^{2}}-\frac{2}{3} \ln \left[2.27 \times 10^{-14} B\right]-\frac{4.4 \times 10^{13}\left(0.855+\ln \left[2.27 \times 10^{-14} B\right]\right)}{B}}{274 \pi}\right)\right)\left.sin \left[\theta_{B}\right]^{2}\right)^{2} \right)\Bigg/ 
\end{align*}

\begin{align*}
 \left(B^{2} n_{\pm}^{2} \left(1-\frac{8.06 \times 10^{7} n_{\pm}}{\nu^{2}}-\frac{1.272-\frac{1.356 \times 10^{27}}{B^{2}}-1.51515 \times 10^{-14} B-\frac{4.4 \times 10^{13}\left(0.307+\ln \left[2.27 \times 10^{-14} B\right]\right)}{B}}{274 \pi}+\right.\right. 
\end{align*}
\begin{align*}
 \left.\left.\left.\left.\frac{1.195-\frac{9.68 \times 10^{26}}{B^{2}}-\frac{2}{3} \ln \left[2.27 \times 10^{-14} B\right]-\frac{4.4 \times 10^{13}\left(0.855+\ln \left[2.27 \times 10^{-14} B\right]\right)}{B}}{274 \pi}\right)^{2}\right)\right)\right)+ 
\end{align*}
\begin{align*}
 \left(2 \pi^{2} \nu^{2}\left(1+\left(4.43415 \times 10^{-15}\left(1-\frac{7.83 \times 10^{12} B^{2}}{\nu^{2}}\right) \nu^{2}\left(\left(\left(\frac{2}{3}-\frac{1.936 \times 10^{27}}{B^{2}}+\frac{4.4 \times 10^{13}\left(0.1447-\ln\left[2.27 \times 10^{-14} B\right]\right)}{B}\right)\right.\right.\right.\right. \right.
\end{align*}
\begin{align*}
 \left(1-\frac{8.06 \times 10^{7} n_{\pm}}{\left(1-\frac{7.83 \times 10^{12} B^{2}}{\nu^{2}}\right) \nu^{2}}+\frac{1.195-\frac{9.68 \times 10^{26}}{B^{2}}-\frac{2}{3} \ln \left[2.27 \times 10^{-14} B\right]-\frac{4.4 \times 10^{13}\left(0.855+\ln \left[2.27 \times 10^{-14} B\right]\right)}{B}}{274 \pi}\right) 
\end{align*}
\begin{align*}
 \left(1-\frac{8.06 \times 10^{7} n_{\pm}}{\nu^{2}}-\frac{1.272-\frac{1.356 \times 10^{27}}{B^{2}}-1.51515 \times 10^{-14} B-\frac{4.4 \times 10^{13}\left(0.307+\ln \left[2.27 \times 10^{-14} B\right]\right)}{B}}{274 \pi}+\right. 
\end{align*}
\begin{align*}
 \left.\left.\frac{1.195-\frac{9.68 \times 10^{26}}{B^{2}}-\frac{2}{3} \ln \left[2.27 \times 10^{-14} B\right]-\frac{4.4 \times 10^{13}\left(0.855+\ln \left[2.27 \times 10^{-14} B\right]\right)}{B}}{274 \pi}\right) \operatorname{sin}\left[\theta_{B}\right]^{2}\right) \Bigg/\Bigg(274 \pi 
\end{align*}
\begin{align*}
 \left.\left(1+\frac{1.195-\frac{9.68 \times 10^{26}}{B^{2}}-\frac{2}{3} \ln \left[2.27 \times 10^{-14} B\right]-\frac{4.4 \times 10^{13}\left(0.855+\ln \left[2.27 \times 10^{-14} B\right]\right)}{B}}{274 \pi}\right)\right)+\left(-1+\frac{5.086 \times 10^{28} B^{2} n_{\pm}^{2}}{\left(1-\frac{7.83 \times 10^{12} B^{2}}{\nu^{2}}\right)^{2} \nu^{6}}+\right. 
\end{align*}
\begin{align*}
 \frac{8.06 \times 10^{7} n_{\pm}}{\left(1-\frac{7.83 \times 10^{12} B^{2}}{\nu^{2}}\right) \nu^{2}}-\frac{1.195-\frac{9.68 \times 10^{26}}{B^{2}}-\frac{2}{3} \ln \left[2.27 \times 10^{-14} B\right]-\frac{4.4 \times 10^{13}\left(0.855+\ln \left[2.27 \times 10^{-14} B\right]\right)}{B}}{274 \pi}+ 
\end{align*}
\begin{align*}
 \left(1-\frac{8.06 \times 10^{7} n_{\pm}}{\left(1-\frac{7.83 \times 10^{12} B^{2}}{\nu^{2}}\right) \nu^{2}}+\frac{1.195-\frac{9.68 \times 10^{26}}{B^{2}}-\frac{2}{3} \ln \left[2.27 \times 10^{-14} B\right]-\frac{4.4 \times 10^{13}\left(0.855+\ln \left[2.27 \times 10^{-14} B\right]\right)}{B}}{274 \pi}\right) 
\end{align*}
\begin{align*}
 \left(1-\frac{8.06 \times 10^{7} n_{\pm}}{\nu^{2}}-\frac{1.272-\frac{1.356 \times 10^{27}}{B^{2}}-1.51515 \times 10^{-14} B-\frac{4.4 \times 10^{13}\left(0.307+\ln \left[2.27 \times 10^{-14} B\right]\right)}{B}}{274 \pi}+\right. 
\end{align*}
\begin{align*}
\left.\left. \left.\left.\frac{1.195-\frac{9.68 \times 10^{26}}{B^{2}}-\frac{2}{3} \ln \left[2.27 \times 10^{-14} B\right]-\frac{4.4 \times 10^{13}\left(0.855+\ln \left[2.27 \times 10^{-14} B\right]\right)}{B}}{274 \pi}\right) \right)\operatorname{sin}\left[\theta_{B}\right]^{2}\right)\right)\Bigg / 
\end{align*}
\begin{align*}
 \left(n_{\pm} \sqrt{\frac{B^{2}}{\nu^{2}}}\left(1-\frac{8.06 \times 10^{7} n_{\pm}}{\nu^{2}}-\frac{1.272-\frac{1.356 \times 10^{27}}{B^{2}}-1.51515 \times 10^{-14} B-\frac{4.4 \times 10^{13}\left(0.307+\ln \left[2.27 \times 10^{-14} B\right]\right)}{B}+}{274 \pi}\right.\right. 
\end{align*}
\begin{align*}
 \left.\left.\left.\left.\frac{1.195-\frac{9.68 \times 10^{26}}{B^{2}}-\frac{2}{3} \ln \left[2.27 \times 10^{-14} B\right]-\frac{4.4 \times 10^{13}\left(0.855+\ln \left[2.27 \times 10^{-14} B\right]\right)}{B}}{274\pi} \right)\right)\right)^{2}\right)\Bigg/
\end{align*}
\begin{align*}
 \left(\left(9.34 \times 10^{-19} B^{4}+\left(1.76 \times 10^{7} B+2 \pi \nu\right)^{2}\right)\left(1+\left(1.96617 \times 10^{-29}\left(1-\frac{7.83 \times 10^{12} B^{2}}{\nu^{2}}\right)^{2} \nu^{6} \operatorname{sec}\left[\theta_{B}\right]^{2}\right.\right.\right. 
\end{align*}
\begin{align*}
 \left(\left(\left(\frac{2}{3}-\frac{1.936 \times 10^{27}}{B^{2}}+\frac{4.4 \times 10^{13}\left(0.1447-\ln \left[2.27 \times 10^{-14} B\right]\right)}{B}\right)\right.\right. 
\end{align*}
\begin{align*}
 \left(1-\frac{8.06 \times 10^{7} n_{\pm}}{\left(1-\frac{7.83 \times 10^{12} B^{2}}{\nu^{2}}\right) \nu^{2}}+\frac{1.195-\frac{9.68 \times 10^{26}}{B^{2}}-\frac{2}{3} \ln \left[2.27 \times 10^{-14} B\right]-\frac{4.4 \times 10^{13}\left(0.855+\ln \left[2.27 \times 10^{-14} B\right]\right)}{B}}{274 \pi}\right) 
\end{align*}
\begin{align*}
 \left(1-\frac{8.06 \times 10^{7} n_{\pm}}{\nu^{2}}-\frac{1.272-\frac{1.356 \times 10^{27}}{B^{2}}-1.51515 \times 10^{-14} B-\frac{4.4 \times 10^{13}\left(0.307+\ln \left[2.27 \times 10^{-14} B\right]\right)}{B}}{274 \pi}+\right. 
\end{align*}
\begin{align*}
 \left.\left.\frac{1.195-\frac{9.68 \times 10^{26}}{B^{2}}-\frac{2}{3} \ln \left[2.27 \times 10^{-14} B\right]-\frac{4.4 \times 10^{13}\left(0.855+\ln \left[2.27 \times 10^{-14} B\right]\right)}{B}}{274 \pi}\right) \operatorname{sin}\left[\theta_{B}\right]^{2}\right)\Bigg/\Bigg(274\pi
\end{align*}
\begin{align*}
 \left.\left(1+\frac{1.195-\frac{9.68 \times 10^{26}}{B^{2}}-\frac{2}{3} \ln \left[2.27 \times 10^{-14} B\right]-\frac{4.4 \times 10^{13}\left(0.855+\ln \left[2.27 \times 10^{-14} B\right]\right)}{B}}{274 \pi}\right)\right)+\left(-1+\frac{5.086 \times 10^{28} B^{2} n_{\pm}^{2}}{\left(1-\frac{7.83 \times 10^{12} B^{2}}{\nu^{2}}\right)^{2} \nu^{6}}+\right. 
\end{align*}
\begin{align*}
 \frac{8.06 \times 10^{7} n_{\pm}}{\left(1-\frac{7.83 \times 10^{12} B^{2}}{\nu^{2}}\right) \nu^{2}}-\frac{1.195-\frac{9.68 \times 10^{26}}{B^{2}}-\frac{2}{3} \ln \left[2.27 \times 10^{-14} B\right]-\frac{4.4 \times 10^{13}\left(0.855+\ln \left[2.27 \times 10^{-14} B\right]\right)}{B}}{274 \pi}+ 
\end{align*}
\begin{align*}
 \left(1-\frac{8.06 \times 10^{7} n_{\pm}}{\left(1-\frac{7.83 \times 10^{12} B^{2}}{\nu^{2}}\right) \nu^{2}}+\frac{1.195-\frac{9.68 \times 10^{26}}{B^{2}}-\frac{2}{3} \ln \left[2.27 \times 10^{-14} B\right]-\frac{4.4 \times 10^{13}\left(0.855+\ln \left[2.27 \times 10^{-14} B\right]\right)}{B}}{274 \pi}\right) 
\end{align*}
\begin{align*}
 \left(1-\frac{8.06 \times 10^{7} n_{\pm}}{\nu^{2}}-\frac{1.272-\frac{1.356 \times 1 \theta^{27}}{B^{2}}-1.51515 \times 10^{-14} B-\frac{4.4 \times 10^{13}\left(0.307+\ln \left[2.27 \times 10^{-14} B\right]\right)}{B}}{274 \pi}+\right. 
\end{align*}
\begin{align*}
 \left.\left.\left.\left.\frac{1.195-\frac{9.68 \times 10^{26}}{B^{2}}-\frac{2}{3} \ln \left[2.27 \times 10^{-14} B\right]-\frac{4.4 \times 10^{13}\left(0.855+\ln \left[2.27 \times 10^{-14} B\right]\right)}{B}}{274 \pi}\right)\right) \operatorname{sin}\left[\theta_{B}\right]^{2}\right)^{2}\right)\Bigg/
\end{align*}
\begin{align*}
 \left(B^{2} n_{\pm}^{2} \left(1-\frac{8.06 \times 10^{7} n_{\pm}}{\nu^{2}}-\frac{1.272-\frac{1.356 \times 10^{27}}{B^{2}}-1.51515 \times 10^{-14} B-\frac{4.4 \times 10^{13}\left(0.307+\ln \left[2.27 \times 10^{-14} B\right]\right)}{B}}{274 \pi}+\right.\right. 
\end{align*}
\begin{align*}
 \left.\left.\left.\left.\frac{1.195-\frac{9.68 \times 10^{26}}{B^{2}}-\frac{2}{3} \ln \left[2.27 \times 10^{-14} B\right]-\frac{4.4 \times 10^{13}\left(0.855+\ln \left[2.27 \times 10^{-14} B\right]\right)}{B}}{274 \pi}\right)^{2}\right)\right)\right)+
\end{align*}
\begin{align*}
 \left(7.76211 \times 10^{-28}\left(1-\frac{7.83 \times 10^{12} B^{2}}{\nu^{2}}\right)^{2} \nu^{8}\left(\left(\left(\frac{2}{3}-\frac{1.936 \times 10^{27}}{B^{2}}+\frac{4.4 \times 10^{13}\left(0.1447-\ln \left[2.27 \times 10^{-14} B\right]\right)}{B}\right)\right.\right.\right.
\end{align*}
\begin{align*}
 \left(1-\frac{8.06 \times 10^{7} n_{\pm}}{\left(1-\frac{7.83 \times 10^{12} B^{2}}{\nu^{2}}\right) \nu^{2}}+\frac{1.195-\frac{9.68 \times 10^{26}}{B^{2}}-\frac{2}{3} \ln \left[2.27 \times 10^{-14} B\right]-\frac{4.4 \times 10^{13}\left(0.855+\ln \left[2.27 \times 10^{-14} B\right]\right)}{B}}{274 \pi}\right) 
\end{align*}
\begin{align*}
 \left(1-\frac{8.06 \times 10^{7} n_{\pm}}{\nu^{2}}-\frac{1.272-\frac{1.356 \times 10^{27}}{B^{2}}-1.51515 \times 10^{-14} B-\frac{4.4 \times 10^{13}\left(0.307+\ln \left[2.27 \times 10^{-14} B\right]\right)}{B}}{274 \pi}+\right. 
\end{align*}
\begin{align*}
 \left.\left.\frac{1.195-\frac{9.68 \times 10^{26}}{B^{2}}-\frac{2}{3} \ln \left[2.27 \times 10^{-14} B\right]-\frac{4.4 \times 10^{13}\left(0.855+\ln \left[2.27 \times 10^{-14} B\right]\right)}{B}}{274 \pi}\right) \operatorname{sin}\left[\theta_{B}\right]^{2}\right)\Bigg/
\end{align*}
\begin{align*}
 \left(274 \pi\left(1+\frac{1.195-\frac{9.68 \times 10^{26}}{B^{2}}-\frac{2}{3} \ln \left[2.27 \times 10^{-14} B\right]-\frac{4.4 \times 10^{13}\left(0.855+\ln \left[2.27 \times 10^{-14} B\right]\right)}{B}}{274 \pi}\right)\right)+ 
\end{align*}
\begin{align*}
 \left(-1+\frac{5.086 \times 10^{28} B^{2} n_{\pm}^{2}}{\left(1-\frac{7.83 \times 10^{12} B^{2}}{\nu^{2}}\right)^{2} \nu^{6}}+\frac{8.06 \times 10^{7} n_{\pm}}{\left(1-\frac{7.83 \times 10^{12} B^{2}}{\nu^{2}}\right) \nu^{2}}-\right. 
\end{align*}
\begin{align*}
 \frac{1.195-\frac{9.68 \times 10^{26}}{B^{2}}-\frac{2}{3} \ln \left[2.27 \times 10^{-14} B\right]-\frac{4.4 \times 10^{13}\left(0.855+\ln \left[2.27 \times 10^{-14} B\right]\right)}{B}}{274 \pi}+ 
\end{align*}
\begin{align*}
 \left(1-\frac{8.06 \times 10^{7} n_{\pm}}{\left(1-\frac{7.83 \times 10^{12} B^{2}}{\nu^{2}}\right) \nu^{2}}+\frac{1.195-\frac{9.68 \times 10^{26}}{B^{2}}-\frac{2}{3} \ln \left[2.27 \times 10^{-14} B\right]-\frac{4.4 \times 10^{13}\left(0.855+\ln \left[2.27 \times 10^{-14} B\right]\right)}{B}}{274 \pi}\right) 
\end{align*}
\begin{align*}
 \left(1-\frac{8.06 \times 10^{7} n_{\pm}}{\nu^{2}}-\frac{1.272-\frac{1.356 \times 10^{27}}{B^{2}}-1.51515 \times 10^{-14} B-\frac{4.4 \times 10^{13}\left(0.307+\ln \left[2.27 \times 10^{-14} B\right]\right)}{B}}{274 \pi}+\right. 
\end{align*}
\begin{align*}
 \left.\left.\left.\left.\frac{1.195-\frac{9.68 \times 10^{26}}{B^{2}}-\frac{2}{3} \ln \left[2.27 \times 10^{-14} B\right]-\frac{4.4 \times 10^{13}\left(0.855+\ln \left[2.27 \times 10^{-14} B\right]\right)}{B}}{274 \pi}\right)\right) \operatorname{sin}\left[\theta_{B}\right]^{2}\right)^{2} \operatorname{tan}\left[\theta_{B}\right]^{2}\right)\Bigg/\Bigg(B^{2} n_{\pm}^{2} 
\end{align*}
\begin{align*}
 \left(9.34 \times 10^{-19} B^{4}+(2 \pi \nu)^{2}\right)\left(1-\frac{8.06 \times 10^{7} n_{\pm}}{\nu^{2}}-\right. 
\end{align*}
\begin{align*}
 \frac{1.272-\frac{1.356 \times 10^{27}}{B^{2}}-1.51515 \times 10^{-14} B-\frac{4.4 \times 10^{13}\left(0.307+\ln \left[2.27 \times 10^{-14} B\right]\right)}{B}}{274 \pi}+ 
\end{align*}
\begin{align*}
 \left.\frac{1.195-\frac{9.68 \times 10^{26}}{B^{2}}-\frac{2}{3} \ln \left[2.27 \times 10^{-14} B\right]-\frac{4.4 \times 10^{13}\left(0.855+\ln \left[2.27 \times 10^{-14} B\right]\right)}{B}}{274 \pi}\right)^{2} 
\end{align*}
\begin{align*}
 \left(1+\left(1.96617 \times 10^{-29}\left(1-\frac{7.83 \times 10^{12} B^{2}}{\nu^{2}}\right)^{2} \nu^{6} \operatorname{sec}\left[\theta_{B}\right]^{2}\left(\left(\left(\frac{2}{3}-\frac{1.936 \times 10^{27}}{B^{2}}+\frac{4.4 \times 10^{13}\left(0.1447-\ln \left[2.27 \times 10^{-14} B\right]\right)}{B}\right)\right.\right.\right.\right. 
\end{align*}
\begin{align*}
 \left(1-\frac{8.06 \times 10^{7} n_{\pm}}{\left(1-\frac{7.83 \times 10^{12} B^{2}}{\nu^{2}}\right) \nu^{2}}+\frac{1.195-\frac{9.68 \times 10^{26}}{B^{2}}-\frac{2}{3} \ln \left[2.27 \times 10^{-14} B\right]-\frac{4.4 \times 10^{13}\left(0.855+\ln \left[2.27 \times 10^{-14} B\right]\right)}{B}}{274 \pi}\right) 
\end{align*}
\begin{align*}
 \left(1-\frac{8.06 \times 10^{7} n_{\pm}}{\nu^{2}}-\frac{1.272-\frac{1.356 \times 10^{27}}{B^{2}}-1.51515 \times 10^{-14} B-\frac{4.4 \times 10^{13}\left(0.307+\ln \left[2.27 \times 10^{-14} B\right]\right)}{B}}{274 \pi}+\right. 
\end{align*}
\begin{align*}
 \left.\left.\frac{1.195-\frac{9.68 \times 10^{26}}{B^{2}}-\frac{2}{3} \ln \left[2.27 \times 10^{-14} B\right]-\frac{4.4 \times 10^{13}\left(0.855+\ln \left[2.27 \times 10^{-14} B\right]\right)}{B}}{274 \pi}\right)\operatorname{sin}\left[\theta_{B}\right]^{2}\right) \Bigg/\Bigg(274 \pi 
\end{align*}
\begin{align*}
 \left.\left(1+\frac{1.195-\frac{9.68 \times 10^{26}}{B^{2}}-\frac{2}{3} \ln \left[2.27 \times 10^{-14} B\right]-\frac{4.4 \times 10^{13}\left(0.855+\ln \left[2.27 \times 10^{-14} B\right]\right)}{B}}{274 \pi}\right)\right)+\left(-1+\frac{5.086 \times 10^{28} B^{2} n_{\pm}^{2}}{\left(1-\frac{7.83 \times 10^{12} B^{2}}{\nu^{2}}\right)^{2} \nu^{6}}+\right. 
\end{align*}
\begin{align*}
 \frac{8.06 \times 10^{7} n_{\pm}}{\left(1-\frac{7.83 \times 10^{12} B^{2}}{\nu^{2}}\right) \nu^{2}}-\frac{1.195-\frac{9.68 \times 10^{26}}{B^{2}}-\frac{2}{3} \ln \left[2.27 \times 10^{-14} B\right]-\frac{4.4 \times 10^{13}\left(0.855+\ln \left[2.27 \times 10^{-14} B\right]\right)}{B}}{274 \pi}+ 
\end{align*}
\begin{align*}
 \left(1-\frac{8.06 \times 10^{7} n_{\pm}}{\left(1-\frac{7.83 \times 10^{12} B^{2}}{\nu^{2}}\right) \nu^{2}}+\frac{1.195-\frac{9.68 \times 10^{26}}{B^{2}}-\frac{2}{3} \ln \left[2.27 \times 10^{-14} B\right]-\frac{4.4 \times 10^{13}\left(0.855+\ln \left[2.27 \times 10^{-14} B\right]\right)}{B}}{274 \pi}\right) 
\end{align*}
\begin{align*}
 \left(1-\frac{8.06 \times 10^{7} n_{\pm}}{\nu^{2}}-\frac{1.272-\frac{1.356 \times 10^{27}}{B^{2}}-1.51515 \times 10^{-14} B-\frac{4.4 \times 10^{13}\left(0.307+\ln \left[2.27 \times 10^{-14} B\right]\right)}{B}}{274 \pi}+\right. 
\end{align*}
\begin{align*}
 \left.\left.\left.\left.\frac{1.195-\frac{9.68 \times 10^{26}}{B^{2}}-\frac{2}{3} \ln \left[2.27 \times 10^{-14} B\right]-\frac{4.4 \times 10^{13}\left(0.855+\ln \left[2.27 \times 10^{-14} B\right]\right)}{B}}{274 \pi}\right)\right) \operatorname{sin}\left[\theta_{B}\right]^{2}\right)^{2}\right)\Bigg / 
\end{align*}
\begin{align*}
 \left(B^{2} n_{\pm}^{2} \left(1-\frac{8.06 \times 10^{7} n_{\pm}}{\nu^{2}}-\frac{1.272-\frac{1.356 \times 10^{27}}{B^{2}}-1.51515 \times 10^{-14} B-\frac{4.4 \times 10^{13}\left(0.307+\ln \left[2.27 \times 10^{-14} B\right]\right)}{B}}{274 \pi}+\right.\right. 
\end{align*}
\begin{align}
 \left.\left.\left.\left.\left.\frac{1.195-\frac{9.68 \times 10^{26}}{B^{2}}-\frac{2}{3} \ln \left[2.27 \times 10^{-14} B\right]-\frac{4.4 \times 10^{13}\left(0.855+\ln \left[2.27 \times 10^{-14} B\right]\right)}{B}}{274 \pi}\right)^{2}\right)\right)\right)\right).
 \label{eq:ap2}
\end{align}
\end{small}

\clearpage 
	
\bsp	
\label{lastpage}

\end{document}